\documentclass{jpp}
\usepackage{graphicx}
\usepackage[utf8]{inputenc}
\usepackage[T1]{fontenc}
\usepackage{amsmath}
\usepackage{float}
\usepackage[dvipsnames]{xcolor}
\usepackage{siunitx}

\newcommand{\Av}{\mathbf{A}}

\newcommand{\Bv}{\mathbf{B}}

\newcommand{\cd}{\cdot}

\newcommand{\dream}{\textsc{Dream}}
\newcommand{\dd}{\mathrm{d}}

\newcommand{\Erf}{{\rm Erf}}

\newcommand{\Lm}{\Lambda_{m}}
\newcommand{\li}{\ell_{\rm i}}

\newcommand{\na}{\nabla}

\renewcommand{\p}{\partial}

\newcommand{\psit}{\psi_{\rm t}}

\newcommand{\Iw}{I_{\rm wall}}

\newcommand{\psie}{\psi_{\rm edge}}
\newcommand{\psiw}{\psi_{\rm wall}}

\newcommand{\jhot}{j_{\rm hot}}
\newcommand{\jre}{j_{\rm re}}
\newcommand{\johm}{j_{\Omega}}
\newcommand{\jtot}{j_{\rm tot}}

\newcommand{\VpVol}{V'}

\newcommand{\Wp}{W_{\rm p}}

\shorttitle{Runaway dynamics in disruptions with current relaxation}
\shortauthor{I.~Pusztai, M.~Hoppe, and O.~Vallhagen}

\title{Runaway dynamics in tokamak disruptions with current relaxation}

\author{Istv\'{a}n~Pusztai\aff{1}
  \corresp{\email{pusztai@chalmers.se}},
  Mathias~Hoppe\aff{2}, \and   Oskar~Vallhagen\aff{1}}

\affiliation{\aff{1}Department of Physics, Chalmers University of Technology, G\"{o}teborg, SE-41296, Sweden
\aff{2}Swiss Plasma Center, Ecole Polytechnique F\'{e}d\'{e}rale de Lausanne, Lausanne, CH-1015, Switzerland}

\begin{document}

\maketitle

\begin{abstract}
The safe operation of tokamak reactors requires a reliable modeling capability of disruptions, and in particular the spatio-temporal dynamics of associated runaway electron currents. In a disruption, instabilities can break up magnetic surfaces into chaotic field line regions, causing current profile relaxation, as well as a rapid radial transport of heat and particles. Using a mean-field helicity transport model  implemented in the disruption runaway modeling framework {\dream}, we calculate the dynamics of runaway electrons in the presence of current relaxation events. 
In scenarios where flux surfaces remain intact in parts of the plasma, a skin current is induced at the boundary of the intact magnetic field region. This skin current region becomes an important center concerning the subsequent dynamics: It may turn into a hot ohmic current channel, or a sizable radially localized runaway beam, depending on the heat transport. If the intact region is in the plasma edge, runaway generation in the counter-current direction can occur, which may develop into a sizable reverse runaway beam. Even when the current relaxation extends to the entire plasma, the final runaway current density profile can be significantly affected, as the induced electric field is reduced in the core and increased in the edge, thereby shifting the center of runaway generation towards the edge.

\end{abstract}

\section{Introduction}
\label{sec:intro}

Recent progress in fusion science has increased confidence in that the energy confinement requirements of up-coming experiments aiming to demonstrate a scientific break-even of the power balance, such as ITER and SPARC, can be met during normal operation \citep{SPARCconf_2020}. However, off-normal events leading to the rapid loss of the energy content of the plasma, called \emph{disruptions} \citep{Hender_2007,BoozerDisr,Lehnen2015}, and associated generation of highly energetic \emph{runaway electron} beams, represent an outstanding challenge of the otherwise most successful tokamak concept for magnetic confinement. 

In a disruption, the thermal energy of the plasma is lost during the rapid \emph{thermal quench}, followed by a resistive decay of the ohmic plasma current, called \emph{current quench}. Parallel electric fields, well in excess of the Connor-Hastie critical electric field \citep{Connor_Hastie} for runaway generation, may then be induced, generating a runaway electron seed \citep{Dreicer1,Harvey2000,MartinSolis2017} which can, in a collisional avalanche process, exponentiate into a macroscopic runaway current \citep{sokolov1979,Rosenbluth_1997,Hesslow_2019}. Owing to the exponential sensitivity of the avalanche to the initial plasma current, high-current reactor relevant devices are predicted to be prone to convert a large fraction of their ohmic current into runaway electron current \citep{BoozerDisr,Breizman_2019,Vallhagen2020}, thereby posing a major threat to the structural integrity of the plasma facing components.
Methods to mitigate the effects of a disruption \citep{Lehnen2015} focus on limiting the exposure of the wall to localized heat losses and to the impact of high current runaway electron beams, as well as on avoiding excessive forces on structural elements. The currently promoted disruption mitigation methods employ a rapid massive material injection, such as through shattered pellet injection \citep{Commaux_2016,Jachmich_2021,Vallhagen2022}, or magnetic perturbations \citep{Tinguely_2021}, to tailor the spatio-temporal properties of dissipating the thermal and magnetic energy content of the plasma. 

Disruption is a strongly nonlinear multi-physics process, with wide ranges of temporal and spatial scales, and kinetic physics at play. Thus progress in developing robust disruption mitigation strategies must inevitably rely on employing a broad range of numerical tools in terms of the trade-off between physical accuracy and numerical cost; these include three-dimensional (3D) magnetohydrodynamic (MHD) simulation codes \citep{Hoelzl2021JOREK,Sovinec2004,Ferraro2016}, flux surface averaged transport solvers \citep{Linder2020}, local kinetic models \citep{Stahl_2016}, sometimes a combination thereof \citep{hoppeASDEX}. The thermal quench often starts with an instability that breaks up the confining magnetic surfaces, leading to a highly elevated transport along multiple channels \citep{Hender_2007}, including electron heat losses, transport of energetic particles, and a redistribution of current density. This is a very consequential process concerning the rest of the disruption, while it is perhaps the most difficult phase of the disruption to model. Models that compromise on spatial dimensionality cannot self-consistently describe this process, while even in 3D MHD simulations it is a challenge \citep{Nardon_2021,boozer_Prevalent} to resolve the process at physically low resistivity.   

A key mechanism of the destruction of flux surfaces during the thermal quench is \emph{fast magnetic reconnection} \citep{BoozerFast}; a rapid topological change of the magnetic field occurring on an ideal MHD timescale, independently of resistivity. During this process, field lines connect regions initially belonging to different magnetic surfaces with different current density and pressure, and the variation of $j_\|/B$ (parallel current density over magnetic field strength) along the field line is relaxed, mediated by the propagation of shear Alfv\'{e}n waves. The effects of such current relaxation events---which do not necessarily extend over the entire plasma volume---are experimentally observed through a sizable temporary increase of the total plasma current $I_p$  \citep{Wesson_1990,de_Vries_2015}, as well as the reduction of the internal inductance of the plasma $\li=\Wp/(\mu_0 R_0 I_p^2/4)$, such that the magnetic energy in the poloidal magnetic field $W_p$ is only slightly reduced. Here $\mu_0$ is the vacuum permeability and $R_0$ is the major radius at the magnetic axis of the unperturbed magnetic field. The low resistivity and the ideal MHD timescale imply that the process approximately conserves magnetic helicity \citep{TaylorRelax,Berger_1999}, a topological measure of magnetic field linkage that can only change on global resistive timescales. Magnetic helicity $H^{\rm M}$ is the volume integral of $\Av\cd\Bv$ with the magnetic vector potential and field, $\Av$ and $\Bv$, which can be expressed in tokamak geometry as $H^{\rm M}=-2\int \psi \, d \psi_t$, with the poloidal and toroidal magnetic fluxes, $\psi$ and $\psi_t$, defined as in Fig.~2 of \citep{Boozer_2017}, and the entire plasma being the integration domain. 

The current density evolution during a thermal quench is too rapid to be experimentally directly observable; in practice the total plasma current and---in case of shaped plasmas---the internal inductance are the quantities that can be experimentally obtained during a disruption. Thus, for practical reasons, several disruption runaway electron studies use the pre-disruption current profile \citep{Linder2020} or assume a flat electric field profile \citep{hoppeASDEX,Klara_2021} as their initial condition, which is then evolved in an inductive-resistive fashion. However, the rearrangement of the current density in the fast magnetic reconnection event, and in particular the development of skin currents at boundaries between chaotic field line regions and intact flux surfaces \citep{Boozer_2017}, is expected to have a significant effect on the development of the runaway current. It has already been shown that the transport of runaway electrons---also caused by the chaotic magnetic field---has a major impact on the evolution of the runaway current density \citep{Svenningsson2021}, but studies concerning the effect of the current relaxation are lacking.         
To capture current relaxation through fast magnetic reconnection in a model that only retains the radial (i.e.~across the unperturbed magnetic flux surfaces) variations, one may employ the mean field model of \citet{Boozer_2018}, that describes the current redistribution as a hyper-diffusion of the poloidal flux, and is constructed to conserve magnetic helicity. 

In this paper we study the effect of current relaxation in the thermal quench phase of tokamak disruptions on the evolution of the ohmic and runaway current components. The plasma parameters considered are ITER-like, and deuterium and neon are assumed to be injected as part of the disruption mitigation. We consider both full-radius and partial reconnection events, with a particular focus on the fate of skin current regions in the latter case. We employ the recently developed \dream{} code \citep{dreampaper}, equipped with an implementation of the helicity transport model of \citet{Boozer_2018}, which allows the self-consistent simulations of plasma cooling and associated runaway electron (RE) dynamics during disruptions. We find that full-radius reconnection tends to shift the center of the runaway generation radially outward. Possible outcomes include a reduced core localized runaway current, or significantly increased runaway currents with hollow radial profile. We also show that skin current regions, when developed, play a central role in the subsequent current evolution, and could turn into long-lived hot ohmic channels or runaway hot-spots. As a possible, somewhat exotic, outcome, ``reverse'' (counter-current) runaway beams may develop in intact edge regions.

The rest of the paper is organized as follows. First, in Sec.~\ref{sec:numerical}, we provide the equations relevant for the current evolution in the numerical model, then in Sec.~\ref{sec:setup} we describe the baseline simulation setup. We first consider full-radius current relaxations; the possible outcomes---core localized and edge localized runaway generation---are discussed in Secs.~\ref{sec:fullradius} and \ref{sec:fullradiusAlt}, respectively. The role of heat transport concerning the fate of a core localized skin current region is discussed in Sec.~\ref{sec:IC}. Finally, we consider the possibility of an intact edge, and associated reverse runaway generation in Sec.~\ref{sec:IE}. Limitations and future directions are touched upon in Sec.~\ref{sec:discussion}, and the conclusions are summarized in Sec.~\ref{sec:conclusions}. 
 
\section{Methods}
\label{sec:methods}
\subsection{Numerical approach}
\label{sec:numerical}
For our simulations we use the \dream{} tool, which is a finite-volume fluid-kinetic framework developed to model runaway electron dynamics in disruptions. For a detailed description of the code we  refer the reader to  \citep{dreampaper}, while here we overview some aspects particularly relevant for our analysis. The code resolves one spatial dimension---a radial coordinate $r$---and it can resolve the entire gyro-averaged momentum space of electrons, or parts of it, parametrized by the magnitude of the momentum $p$ and $\xi=p_\|/p$, with the component of the momentum along the magnetic field $p_\|$  (taken at the lowest magnetic field position along a collisionless orbit). It has various reduced models for the momentum-space dynamics as well; here we will predominantly use a fluid model that retains only a thermal bulk, characterized by a density $n_e$, a temperature $T_e$, and an ohmic current density $\johm$; and a runaway electron population, characterized by the current density they carry $\jre$. In the fluid model it is assumed that the runaways, which have the density $ n_{\rm re}$, move with the speed of light $c$ parallel to the magnetic field, hence $\jre =e n_{\rm re} c$, where $e$ is the elementary charge. 

The total current density is computed through evolving the poloidal flux  
\begin{equation}
    \frac{\p \psi}{\p t}=-\mathcal{R}+\mu_0\frac{\p}{\p\psit}\left( \psit \Lm \frac{\p}{\p \psit}\frac{\jtot}{B}\right);
    \label{psievol}
\end{equation}
in this equation the toroidal flux $\psit(r)=(2\pi)^{-1}\int_0^r \VpVol \langle \Bv \cd \na \varphi\rangle\, \dd r'$ is used as a radial coordinate, with $\langle\cdot\rangle$ denoting flux surface average, $V'$ is the spatial Jacobian (incremental volume enclosed by two infinitesimally close flux surfaces), and  $\mathcal{R}=2 \pi \langle\mathbf{E}\cdot\mathbf{B}\rangle/\langle\mathbf{B}\cdot\nabla\varphi\rangle$, with the toroidal angle $\varphi$. This flux function term is equal to the loop voltage when the second term in the right hand side of (\ref{psievol}) vanishes, and it describes ohmic dissipation, as we may express the term $\langle\mathbf{E}\cdot\mathbf{B}\rangle=\johm \langle B^2\rangle/(\sigma B)$, with $\sigma$ the electric conductivity, and $\johm$ the ohmic current density\footnote{In kinetic simulations $\johm$ is replaced by $\johm-\delta j_{\rm corr}$, with a correction that plays a role  in the presence of rapid time variation, as explained at Eq.~(29) of \citep{dreampaper}.}. The total current density $\jtot$ includes the ohmic and the runaway current densities,  $\johm$ and  $\jre$, and in some kinetic simulations $\jhot$ as well, the current density carried by electrons at an intermediate energy range between the thermal bulk and the highly energetic runaways. The total current density is related to the poloidal flux through
\begin{equation}
    2\pi \mu_0 \langle \Bv\cd \na \varphi \rangle \frac{\jtot}{B}=\frac{1}{\VpVol} \frac{\partial}{\partial r}\left[ \VpVol \left\langle \frac{|\nabla r|^2}{R^2}\right\rangle  \frac{\partial \psi}{\partial r}   \right],
    \label{eqjtot}
\end{equation}
where $R$ is the major radius of a given point. From the form of (\ref{eqjtot}) we can see that the term involving $\Lm$ in Eq.~(\ref{psievol}) includes a $4^{\rm th}$ order radial derivative of $\psi$, and as such it describes a hyper-diffusion of $\psi$ wherever $\Lm$ is non-zero. This term describes a local transport of magnetic helicity, such that it conserves the total magnetic helicity. That the form of Eq.~(\ref{psievol}) is constructed to respect this conservation property is most easily seen by considering $d_t H^{\rm M}=-2 \int \dd \psi_t\, \partial_t \psi$. Clearly, the contribution of the second term on the r.h.s. of Eq.~(\ref{psievol}) to $\partial_t \psi$---which is of the form $\partial_{\psi_t}(\dots)$---integrates to zero if it vanishes at the boundaries; that is, helicity can only be transported across the plasma boundary, but not created or destroyed inside. The numerical conservation of $H^{\rm M}$ is demonstrated in Appendix~\ref{helicity}. In the simulations, the helicity transport coefficient $\Lm$ can have an arbitrary prescribed spatio-temporal structure.  

Currents in passive structures are taken into account through the boundary condition on Eq.~(\ref{psievol}). The relevant equations are the following
\begin{equation}
\psie \equiv \psi(r=a)= \psiw - M_{\rm ew} I(a),\label{eq:psie}
\end{equation}
\begin{equation}
\psiw \equiv \psi(r=b)= -L_{\rm ext}\left[I(a) + \Iw\right],\label{eq:psiw}
\end{equation}
\begin{equation}
\mathcal{R}^{\rm wall} = R_{\rm wall}\Iw,\label{eq:RwIw}
\end{equation}
where $a$ and $b$ are the minor radii corresponding to the plasma edge and the wall. $I(r)$ is the total plasma current within a given radius $r$. The edge-wall mutual inductance is $M_{\rm ew}=(2\pi)^2\mu_0\int_a^b dr(V'\langle |\nabla r|^2/R^2 \rangle)^{-1}$, which in the cylindrical limit reduces to $\mu_0R_{0} \ln (b/a)$. The external inductance is $L_{\rm ext}=\mu_0R_{0} \ln (R_{0}/b)$, with $R=R_{0}$ at the magnetic axis. The inputs to the wall model are the resistive timescale of the wall $\tau_w=L_{\rm ext}/R_{\rm wall}$, and $b$. 

\subsection{Simulation setup}
\label{sec:setup}
The disruption simulations assume an initially ($t<0$) pure deuterium-tritium plasma  (with even isotope concentrations). 
Specifically, the initial electron density is spatially constant $10^{20}\,\rm m^{-3}$, the temperature is parabolic with $20\,\rm keV$ on-axis, and the current density corresponds to a total plasma current of $15\,\rm MA$. The simulations consider an ITER-like magnetic geometry with $R_0=6\,\rm m$, $a=2\,\rm m$, $b=2.15\, \rm m$, $B(r=0)=5.3\,\rm T$, and a resistive wall time of $\tau_w=0.5\,\rm s$, as well as a Miller model equilibrium \citep{Miller98}, with the radially varying realistic shaping parameters given in Appendix~\ref{sec:initailprof}.

At $t=0$ an instantaneous and homogeneous deposition of additional neutral deuterium and neon is done. At the same time an elevated transport of electron heat and energetic electrons is activated, along with a current profile relaxation, as described by the appropriate transport coefficients, to emulate fast magnetic reconnection and associated break-up of flux surfaces. At $t=6\,\rm ms$, when the maximum plasma temperature has dropped to $\approx 100\,\rm eV$ and the current density has already relaxed, the transport of energetic electrons and magnetic helicity is switched off, and the electron heat transport is strongly reduced. This is to account for the reformation of flux surfaces as the  drive of the instability is removed. A weaker electron heat diffusivity remains active until the end of the simulation. Toroidicity effects are accounted for in the conductivity through the model by \citet{Redl}, and in the runaway generation mechanisms.

The \dream{} simulations are performed in fully fluid mode, unless stated otherwise. The Dreicer runaway generation rate is calculated using a neural network \citep{Neural}, Compton scattering and tritium decay seed sources are accounted for as in \citep{MartinSolis2017,Vallhagen2020}, the hot-tail seed is calculated using the model in Sec.~4.2 of \citep{SvenningssonMsc}, and the avalanche growth rate accounts for partial screening \citep{Hesslow_2019}. 

The bulk electron temperature evolution is calculated from the time dependent energy balance throughout the simulation, according to Eq.~(43) in \citep{dreampaper}, accounting for ohmic heating, line and recombination radiation and bremsstrahlung, as well as a radial heat transport.  Recombined deuterium is assumed to be opaque to Lyman line radiation, which can have a non-negligible effect on the post-thermal quench plasma temperature and indirectly on the avalanche gain \citep{Vallhagen2022}.  Note that here we do not evolve ion temperatures separately and do not have a kinetic runaway population, thus the electron-ion heat exchange term, and the kinetic term describing heating by runaway electrons, are zero. However the latter process is approximately accounted for by a term $j_{\rm re} E_{\rm c}$, with $E_{\rm c}=e^3n_e \ln \Lambda_c/(4\pi \epsilon_0 m_e c^2)$ the critical electric field, $\epsilon_0$ the vacuum permittivity, and $m_e$ the electron mass.  
In the definition of $E_{\rm c}$ we take $\ln \Lambda_c$, the momentum dependent Coulomb logarithm, defined in Eq.~(18) of \citep{dreampaper}, at the momentum $p=20\, m_e c$; in addition $\ln \Lambda_c$ depends on the bulk electron density and temperature, $n_e$ and $T_e$.

In the early phase of the thermal quench electron heat losses are dominated by a transport along the chaotic magnetic field lines. To capture this, we use a Rechester-Rosenbluth-type model in the collisionless limit \citep{RechRos}, with a heat diffusivity given by $D_W\approx 2\sqrt{\pi} R_0 v_{te} (\delta B/B)^2$, where  $v_{te}=\sqrt{2 T_e/m_e}$ is the local electron thermal speed, and $\delta B/B$ is the relative magnetic perturbation amplitude\footnote{This approximate expression for $D_W$ represents the large aspect ratio and non-relativistic temperature limit of the expression used in \dream{}, given by Eq.~(B.5) of \citep{dreampaper}.}. This transport coefficient is applied spatially homogeneously even in the cases where part of the plasma is assumed to have intact magnetic surfaces (where no current relaxation and runaway transport is active). This is to be able to reach sufficiently low temperatures to initiate the radiative temperature collapse without the need to modify the impurity content, thereby making it easier to compare various cases. Also, technically the same form of transport coefficient with reduced $\delta B/B$ is applied after the thermal quench, throughout the simulation, while physically electron heat transport might then be caused by electrostatic turbulence. 

During the thermal quench we also account for a diffusive transport of runaway electrons using a diffusion coefficient of similar form but with parallel streaming along the perturbed field lines at the speed of light $D_{\rm RE}=\pi R_0 c (\delta B/B)^2$. We employ here the same $\delta B/B$ as for electron heat transport, but only in the chaotic field line regions. 
 
The simulations use $200$ radial grid cells, and in cases with skin current regions, $150$ of these are packed around the radial region with sharp current density variation. The first microsecond of the simulation, when the injected neutral material ionizes, is resolved by $1000$ time steps. The rest of the simulation typically uses a time step of $\Delta t= 6.6\,\rm \mu s$.

\section{Results}
\label{sec:results}
\subsection{Full-radius current relaxation; core localized runaways}
\label{sec:fullradius}

First we briefly discuss the temperature evolution through our baseline case; in most cases a similar qualitative behavior is observed, with the exception of cases with strong ohmic skin currents, to be discussed in Sec.~\ref{sec:IC}.  After the instantaneous and homogeneous deposition of neutral deuterium and neon of densities $n_{\rm D,inj}=7\cdot 10^{20}\,\rm m^{-3}$ and $n_{\rm Ne,inj}=1.5\cdot 10^{19}\,\rm m^{-3}$, the plasma temperature drops on a timescale of $0.01$-$0.1\,\rm \si{\micro\second}$ as these species get ionized and the plasma diluted. The peak temperature drops from $20$ to $\sim 2\, \rm keV$. Then the electron heat transport corresponding to a magnetic perturbation amplitude of $\delta B/B = 3.5\cdot 10^{-3}$ cools the plasma further on a $\rm ms$ time scale\footnote{It is in fact this characteristic cooling timescale that was targeted by choosing this specific value of $\delta B/B$.}, until $T_e$ reaches $\sim 100\,\rm eV$, at which point radiative energy losses become dominant and cool the plasma further to the $1$-$10\,\rm eV$ range very rapidly. The injected densities are chosen to produce a disruption in the parameter region deemed favorable in \citep{vallhagen_2020}, producing relatively low maximum runaway currents, and current quench time scales within tolerable limits. 

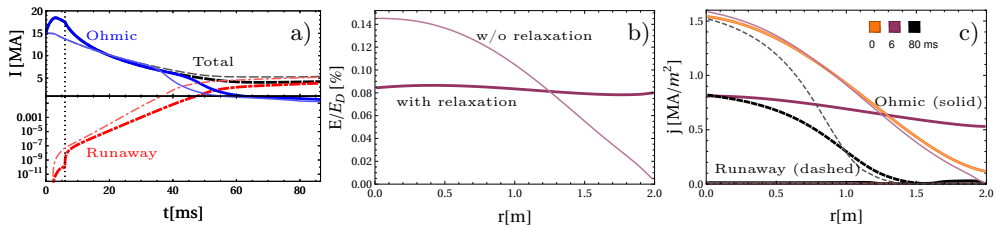
\begin{figure}
    \centering
    \includegraphics[width=0.325\textwidth]{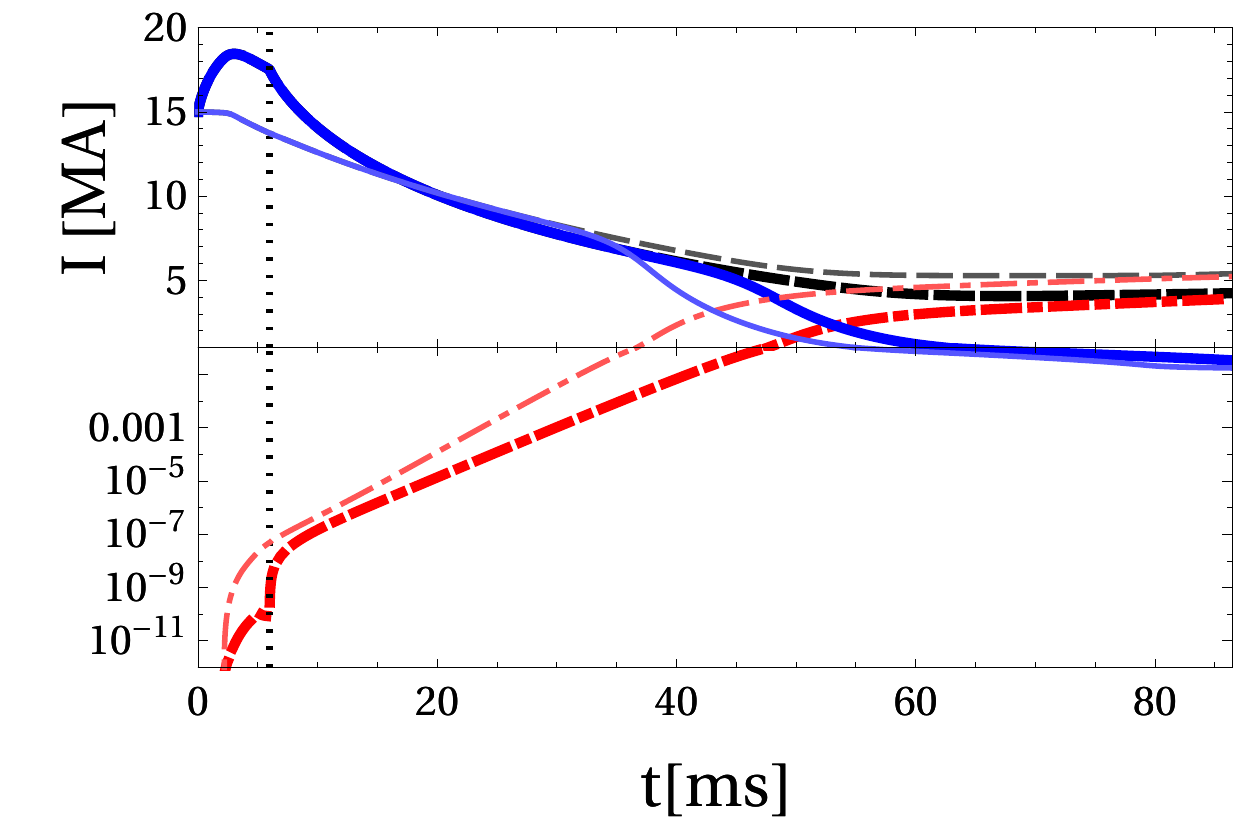}
    \put(-13,70){a)}
    \put(-90,70){\tiny  \textcolor{blue}{Ohmic}}
    \put(-90,26){\tiny  \textcolor{red}{Runaway}}
    \put(-50,60){\tiny  \textcolor{black}{Total}}
    \includegraphics[width=0.33\textwidth]{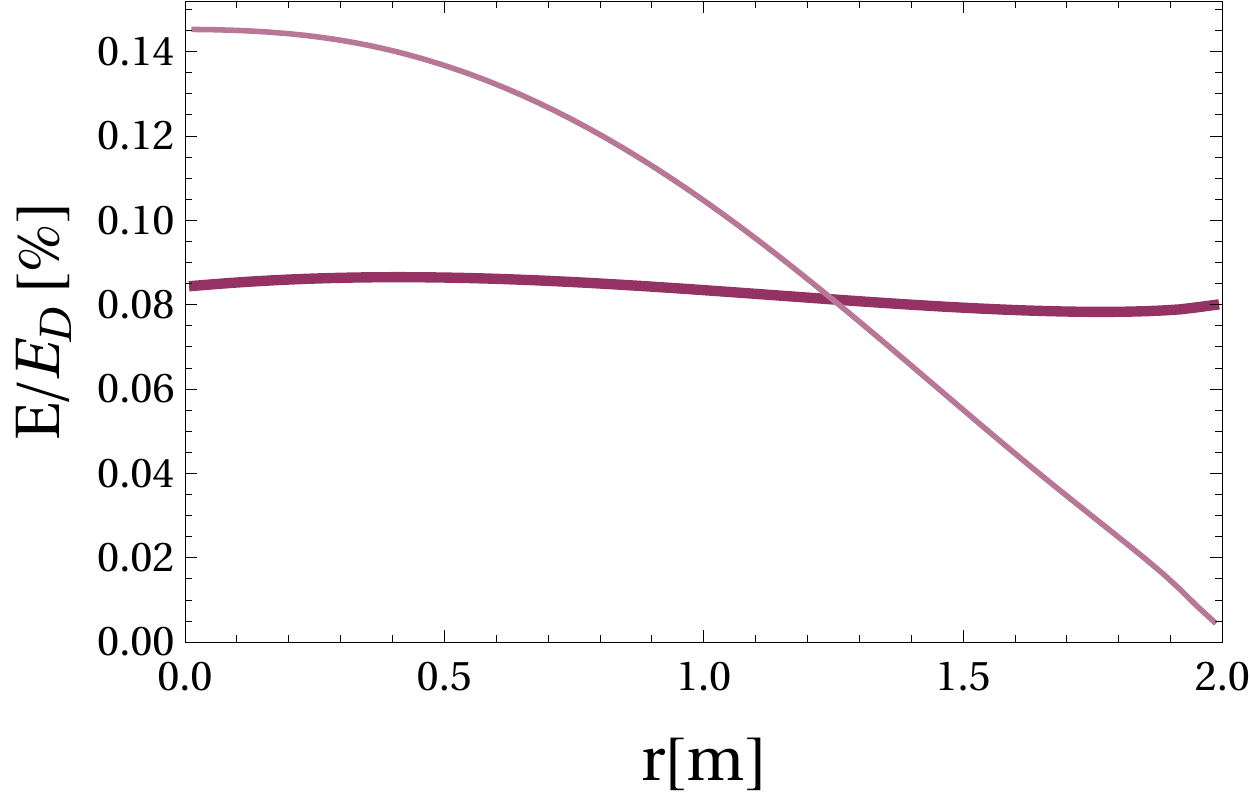}
    \put(-13,70){b)}
    \put(-70,70){\tiny  \textcolor{black}{w/o relaxation}}
    \put(-100,45){\tiny  \textcolor{black}{with relaxation}}
    \includegraphics[width=0.325\textwidth]{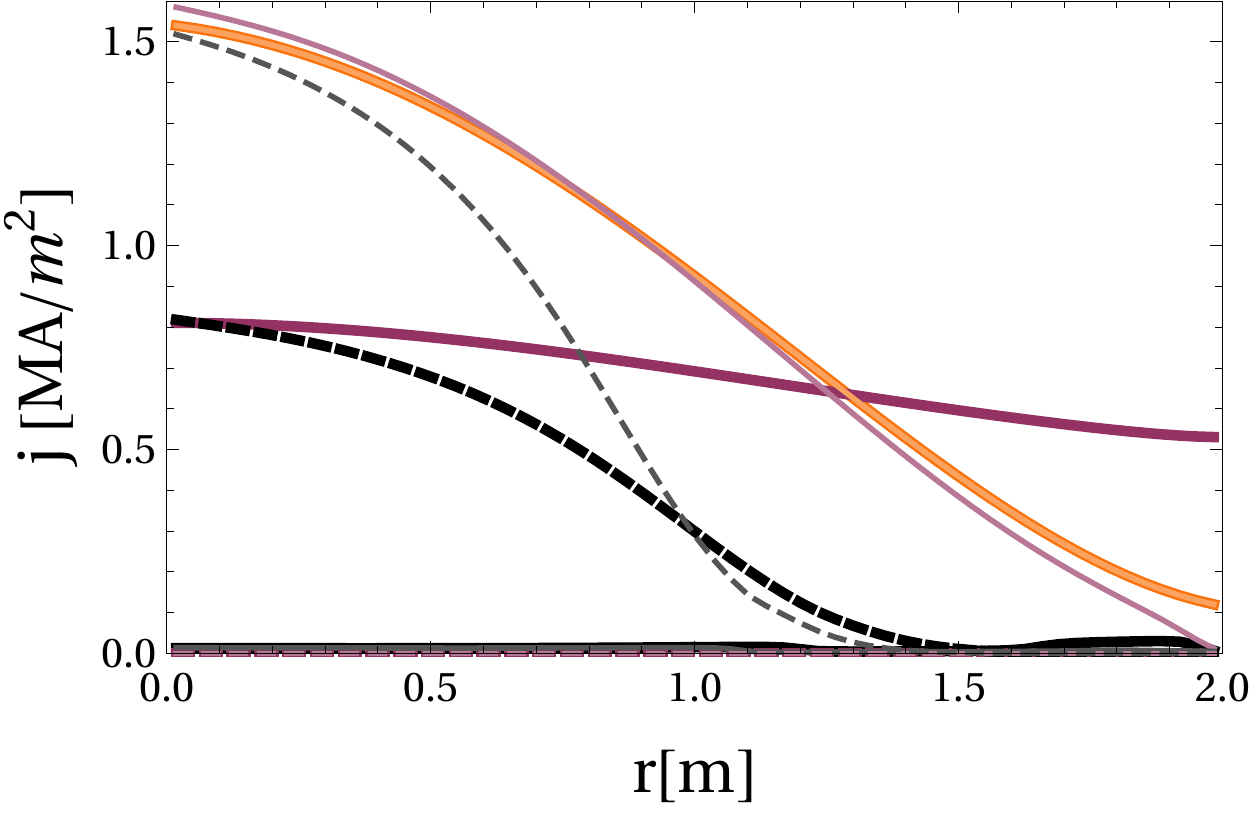}
    \put(-13,70){c)}
    \put(-50,65){ \includegraphics[scale=0.45]{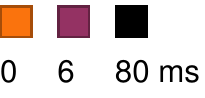}}
    \put(-105,20){\tiny  \textcolor{black}{Runaway (dashed)}}
    \put(-45,45){\tiny  \textcolor{black}{Ohmic (solid)}}
    \caption{Comparison of simulations without (thin light curves) and with (thick dark curves) a full-radius current relaxation. The latter case employs spatially homogeneous helicity and runaway electron transport coefficients in the first  $6\,\rm ms$ (marked by dotted vertical line in Panel a) Panel a: Time evolution of the total plasma current (dashed) and its ohmic (solid) and runaway (dash-dotted) components.  Panel b: Radial profiles of the electric field normalized to the local Dreicer field at $t=6\,\rm ms$. Panel c: Radial profiles of the current density, with ohmic (solid) and runaway (dashed) components. Profiles at $t=\{0,\,6,\, 80\}\rm ms$ are shown with orange, purple and black lines. Note that the runaway component is significant only in the third time point, and the ohmic component in the first two time points.      
    }
    \label{fig:C12_IP}
\end{figure}

Once the thermal quench is complete the electron heat transport is reduced to a level corresponding to $\delta B/B = 4\cdot 10^{-4}$ (at $6\,\rm ms$, indicated by the dashed vertical line in Fig.~\ref{fig:C12_IP}a). Then the temperature evolves on the $\sim 10\,\rm ms$ timescale of the current quench, and it is determined by an approximate balance of ohmic heating (later the friction of energetic electrons on the bulk) and radiative losses.

Without current relaxation (thin light lines in Fig.~\ref{fig:C12_IP}), the total ohmic current (solid line in Fig.~\ref{fig:C12_IP}a) decreases monotonically. After a short period of hot-tail runaway generation that is concluded at $5\,\rm ms$, the dominant seed generation mechanism is tritium decay; while the Dreicer and Compton seeds remain negligible in comparison. Once the runaway current (dash-dotted) reaches $\sim \rm MA$ level through avalanche, the ohmic current drops on the time scale of the avalanche growth, and the runaway current saturates around $5.3\,\rm MA$. Then the RE current keeps increasing only very slowly as poloidal magnetic filed is diffusing back into the vacuum chamber through the resistive wall. 

In case a spatially homogeneous hyperdiffusivity of $\Lambda = 3\cdot 10^{-2} \, \rm Wb^2 m/s$ is applied in the first $6\,\rm ms$ (solid dark lines in Fig.~\ref{fig:C12_IP}), when the flux surfaces are assumed to be broken up, we find that the ohmic current first exhibits a peak of $18.5\,\rm MA$ at $3\,\rm ms$. This is the ``$I_p$-spike'' regularly observed in plasma disruptions, as the current density is radially redistributed at approximately constant magnetic helicity. The ohmic current density indeed flattens\footnote{Note that $j$, the outboard mid-plane value of the parallel current density, does not quite get radially constant, which is partly due to that hyperdiffusion flattens $j/B$, where the outboard magnetic field strength $B$ decreases radially due to toroidicity.}, as seen in Fig.~\ref{fig:C12_IP}c (thick purple line).  

The ohmic current is sustained by the induced electric field, which thus needs to adjust itself to the current redistribution. This means that the electric field must increase at the edge and reduce in the core compared to the case without current relaxation, which is clearly seen in Fig.~\ref{fig:C12_IP}b, showing the electric field normalized to the Dreicer field $E_D=(e^3 n_e \ln \Lambda_c)/(2\pi \epsilon_0^2 m_e v_{te}^2)$. We note that the changes in $E/E_D$ are more affected by differences in $E$ rather than $E_D$, as changes in the temperature profiles are modest. 

That current relaxation can lead to strongly disparate outcomes  concerning the runaway current evolution can be better understood from the observation that the runaway electron density profiles often grow as to replace the ohmic current, thus the initial ohmic current density acts as an envelope to the final runaway current density (even if there are exceptions from this rough rule of thumb, due to trapping, finite wall time and electric field diffusion). Indeed, we find that the final runaway current densities (dashed lines in Fig.~\ref{fig:C12_IP}c) are everywhere lower than the initial ohmic current density. In the case with current relaxation the same observation can be done, but now with the relaxed current profile playing the role of the envelope. As in both cases the runaway density peaks in the core, this translates to a reduced final runaway electron current in the case with current relaxation (from $5.3\,\rm MA$ to $3.9\,\rm MA$). 

In the first $6\,\rm ms $ of the case with current relaxation runaway electron transport is also active, corresponding to $\delta B/B = 3.5\cdot 10^{-3}$ (consistently with the electron heat transport). This reduces the runaway electron seed, as seen in Fig.~\ref{fig:C12_IP}a. However after the flux surfaces are re-formed (following the dotted vertical line) the runaway current growth quickly recovers, and it is not as much due to the reduced seed, rather due to a slightly reduced avalanche growth rate, that the runaway current reaches macroscopic values later in the case with current relaxation.

\subsection{Full-radius current relaxation; edge localized runaways}
\label{sec:fullradiusAlt}

\begin{figure}
    \centering
    \includegraphics[width=0.33\textwidth]{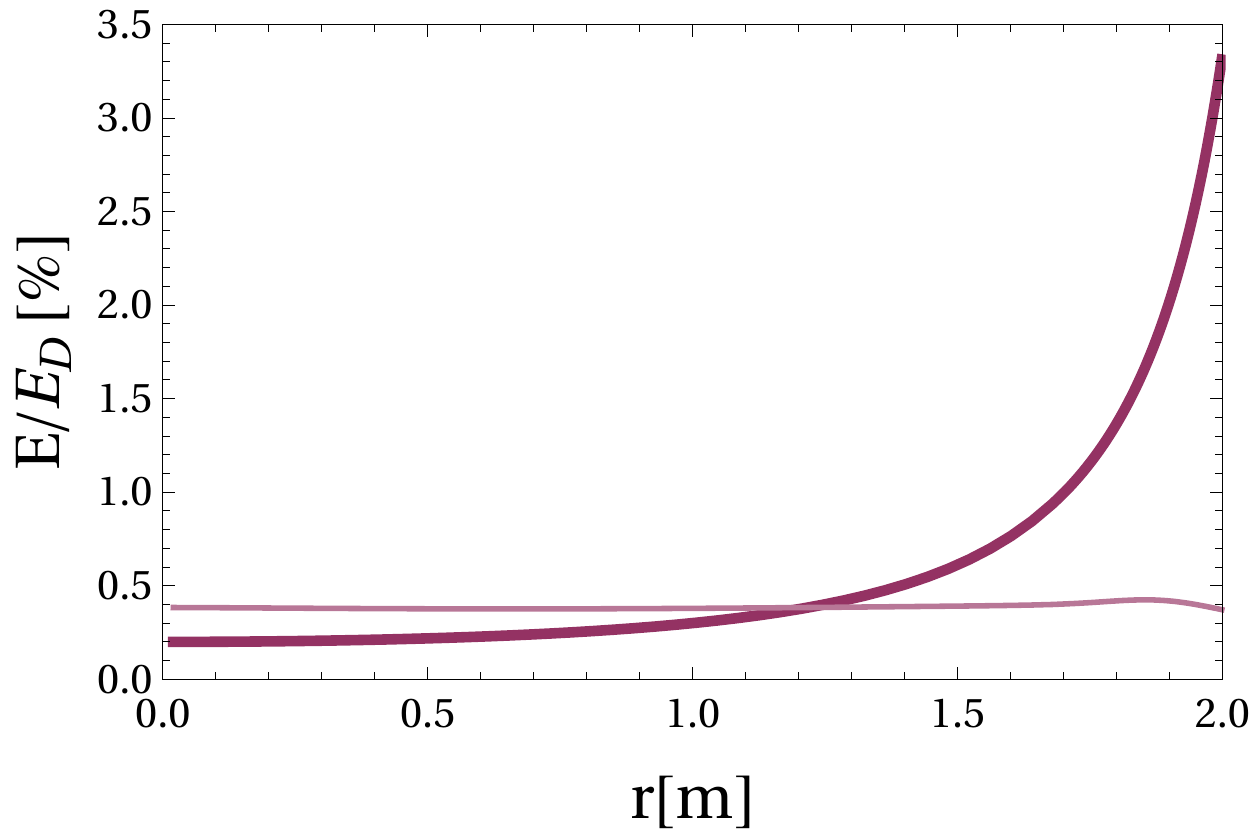}
    \put(-13,70){a)}
    \put(-105,25){\tiny  \textcolor{black}{w/o relaxation}}
    \put(-58,50){\tiny  \textcolor{black}{with relaxation}}
    \includegraphics[width=0.33\textwidth]{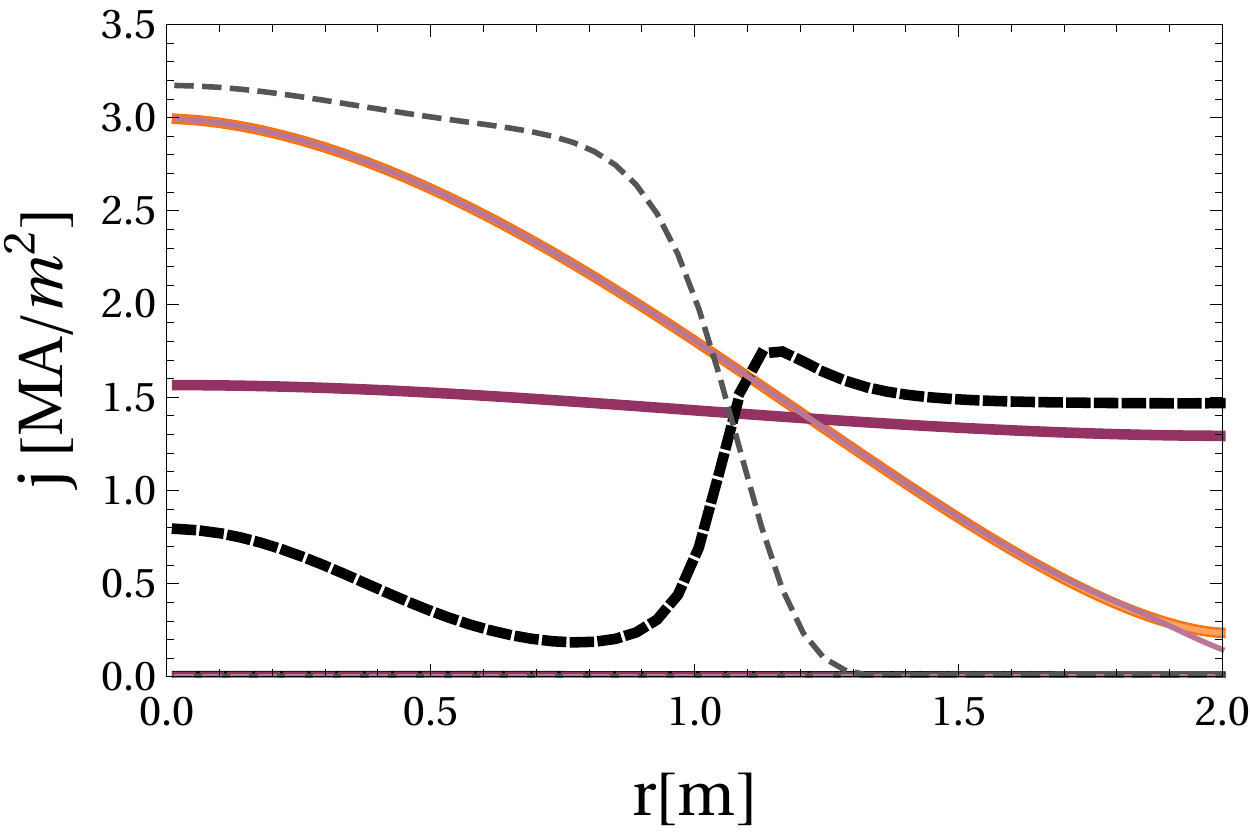}
    \put(-13,70){b)}
    \put(-50,65){ \includegraphics[scale=0.45]{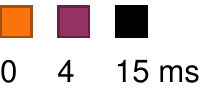}}
    \put(-110,32){\tiny  \textcolor{black}{Runaway (dashed)}}
    \put(-105,46){\tiny  \textcolor{black}{Ohmic (solid)}}
    \caption{Alternative scenario that yields a hollow runaway profile with current relaxation. Panel a: Radial profiles of the electric field normalized to the Dreicer field at $t=4\,\rm ms$. Panel b: current density profiles with ohmic (solid) and runaway (dashed) components. Thinner lines with lighter color correspond to a simulation without current relaxation; thicker darker lines use a spatially homogeneous $\Lm$, applied in the first  $10\,\rm ms$.
    Profiles at $t=\{0,\,4,\, 15\}\rm ms$ are shown with orange, purple and black lines. Note that the runaway component is significant only in the third time point, and the ohmic component in the first two time points.
    }
    \label{fig:alternativeEj}
\end{figure}
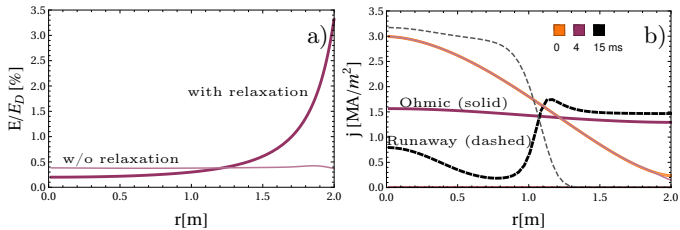

To illustrate that the current flattening could potentially lead to a very different outcome, we consider another case with an ITER-sized plasma (with initial profiles different from those of the baseline). In this case the initial profiles and the simulation setup is such that it favors Dreicer seed generation and the development of a higher electric field in the edge. Such differences include a higher current density, a relatively low post thermal quench temperature and a lower edge electron density. This case has a reduced physics fidelity, and as such, it is likely less representative of the behavior in ITER. For instance, it uses a prescribed temperature variation, the evolution of ion charge states and runaway electron transport are disabled, and no shaping and toroidicity effects are included. More details of corresponding settings are provided in Appendix~\ref{sec:initailprof}. 

Unlike in our baseline case, where $E/E_D$ went from centrally peaked to flat when current relaxation was applied, in this case it goes from mostly flat to peaked towards the edge (compare Fig~\ref{fig:alternativeEj}a to Fig~\ref{fig:C12_IP}b). Note also the significantly higher values of $E/E_D$, resulting in a rapid, and almost full, ohmic-to-runaway current conversion. In the case without current relaxation the final runaway current is peaked in the core, as shown by the thin lines in Fig~\ref{fig:alternativeEj}b, similarly to the baseline. Now  the initial current density does not quite envelope the final runaway current, due to an electric field diffusion into regions where the runaway current is already high. More importantly, when a radially constant $\Lm =1.5\cdot 10^{-2} \,\rm Wb^2 m/s$ is applied in the first $10\,\rm ms$ of the simulation (solid lines), the runaway current profile becomes edge localized and hollow. In fact, the runaway current first grows at the edge, and once it replaces the ohmic current density it continues to grow radially inward. 
Since the weight of the current density towards the total current increases radially, the final total runaway current corresponding to the case with current relaxation is higher than that without current relaxation. This is unlike the situation in our baseline case shown in Fig~\ref{fig:C12_IP}. 

We may conclude then, that a current relaxation that extends over the entire plasma---reducing the electric field in the core and increasing it in the edge---tends to decrease the final runaway current as long as the conditions for runaway generation are more favorable in the core, and increase it if the plasma is more prone to develop edge-localized runaway currents.   

\subsection{Intact core and the role of heat transport}
\label{sec:IC}

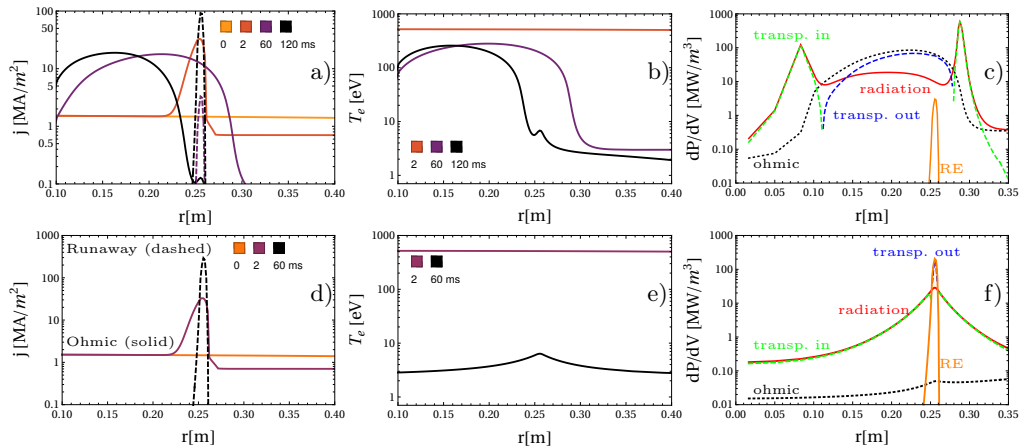
\begin{figure}
    \centering
    \includegraphics[width=0.33\textwidth]{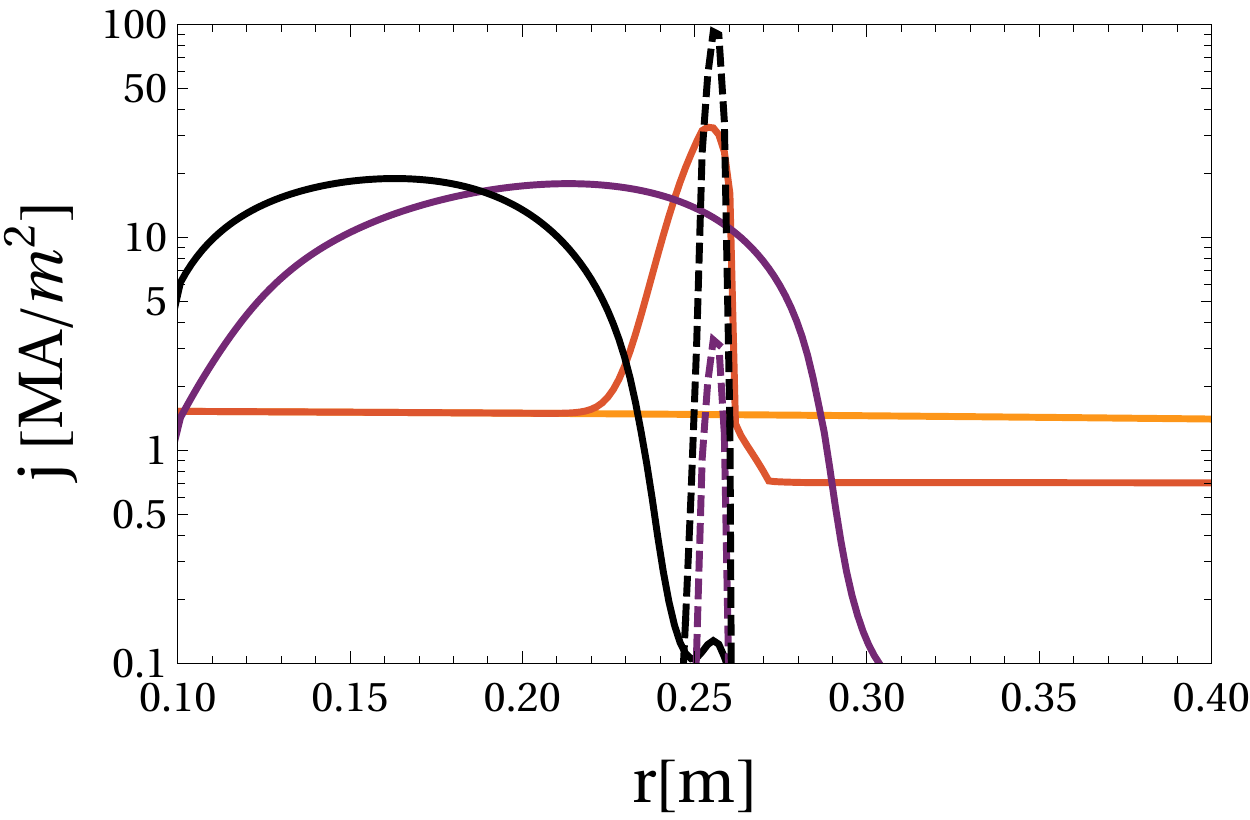}
    \put(-13,55){a)}
    \put(-50,65){ \includegraphics[scale=0.45]{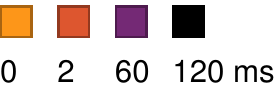}}
    \includegraphics[width=0.33\textwidth]{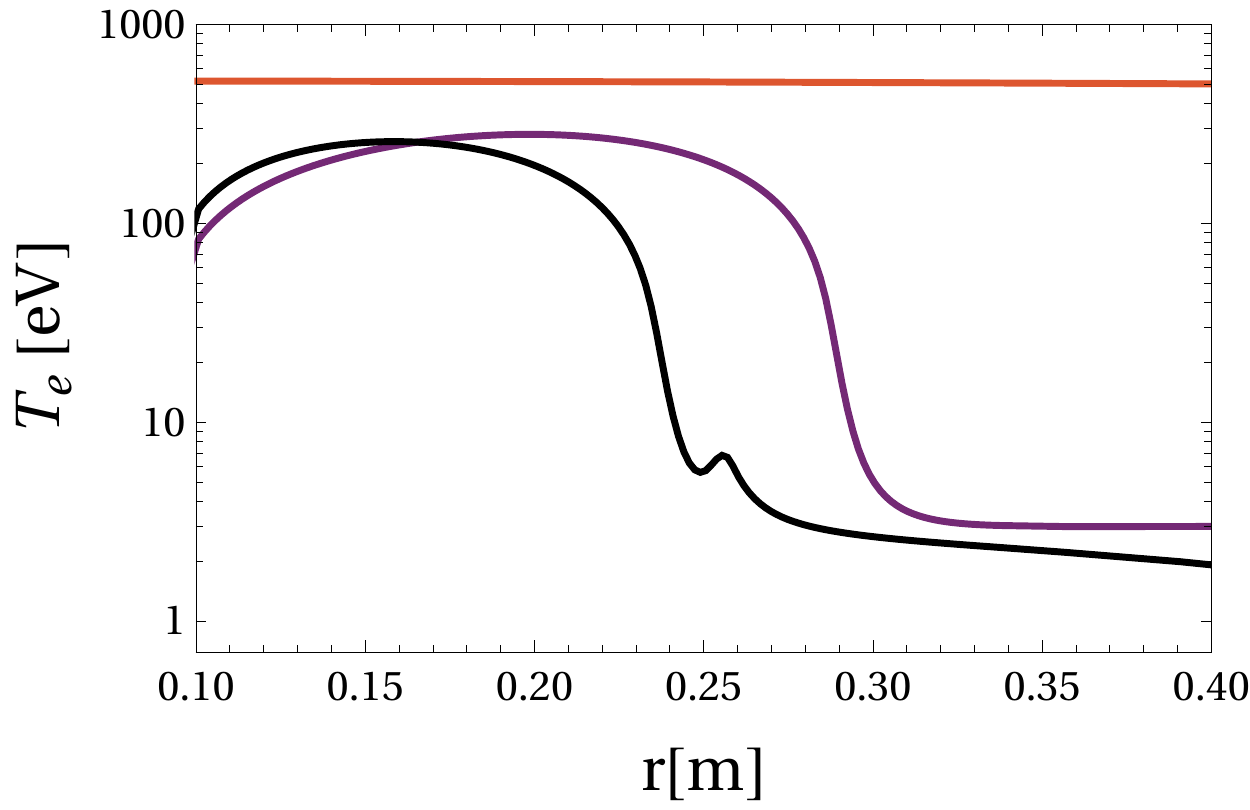}
    \put(-13,55){b)}
    \put(-105,20){ \includegraphics[scale=0.45]{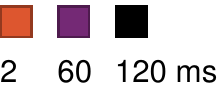}}
    \includegraphics[width=0.33\textwidth]{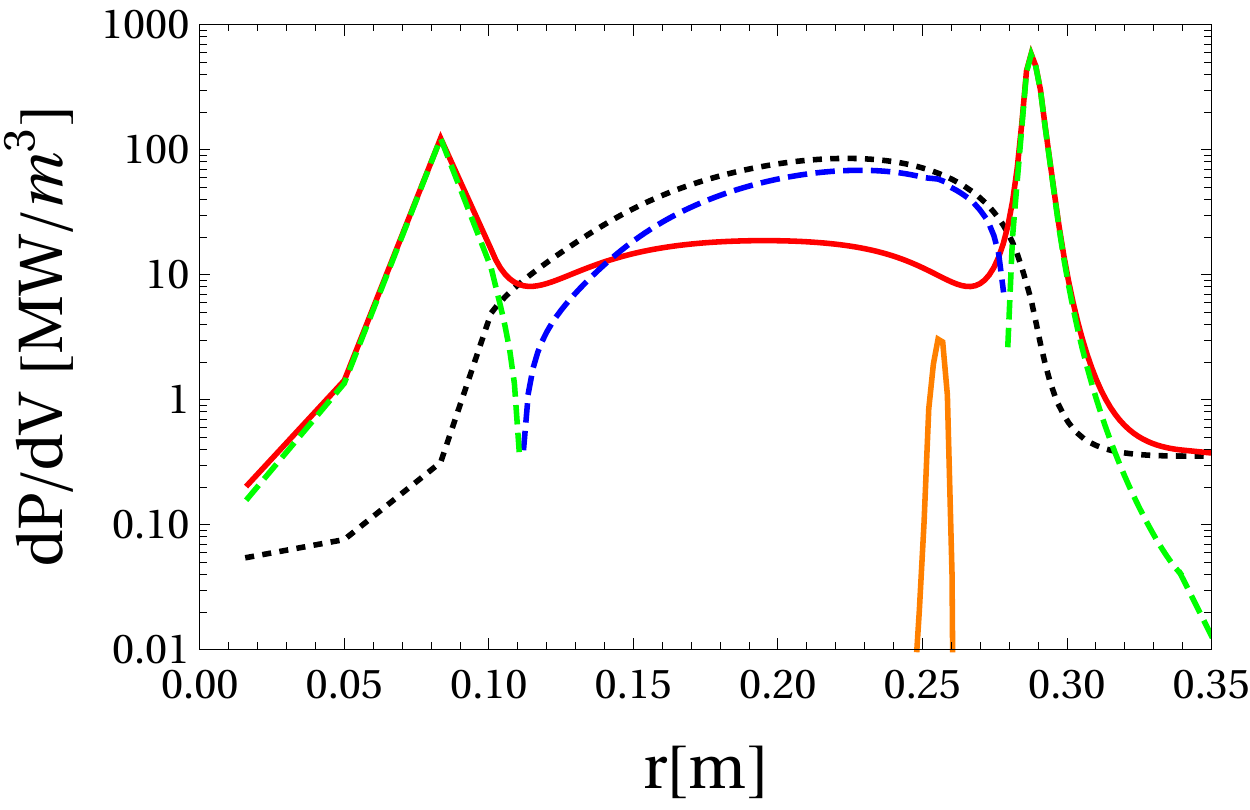}
    \put(-30,20){{\tiny \textcolor{orange}{RE}}}
    \put(-100,22){{\tiny \textcolor{black}{ohmic}}}
    \put(-70,40){{\tiny \textcolor{blue}{transp.~out}}}
    \put(-60,50){{\tiny \textcolor{red}{radiation}}}
    \put(-100,70){{\tiny \textcolor{green}{transp.~in}}}
    \put(-13,55){c)}
    \\
    \includegraphics[width=0.33\textwidth]{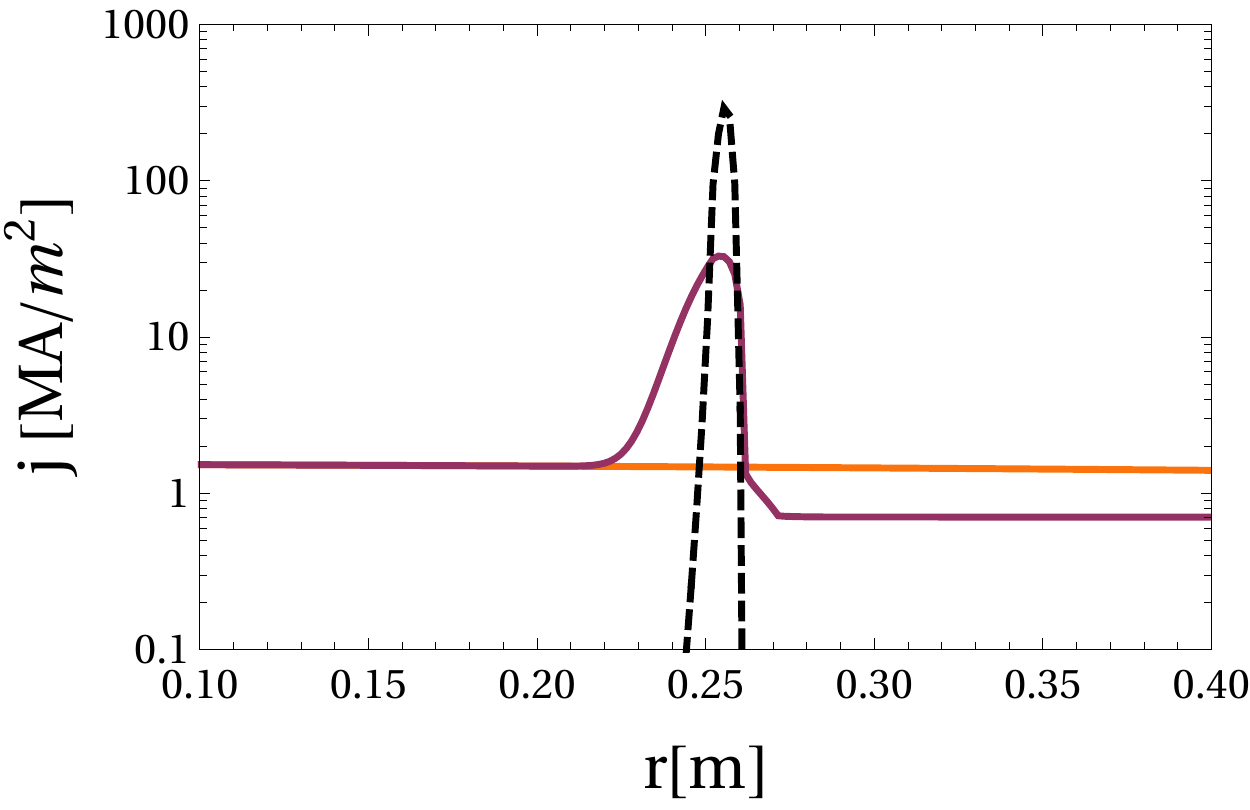}
    \put(-13,55){d)}
    \put(-45,65){ \includegraphics[scale=0.45]{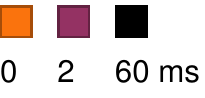}}
    \put(-105,73){\tiny  \textcolor{black}{Runaway (dashed)}}
    \put(-105,38){\tiny  \textcolor{black}{Ohmic (solid)}}
    \includegraphics[width=0.33\textwidth]{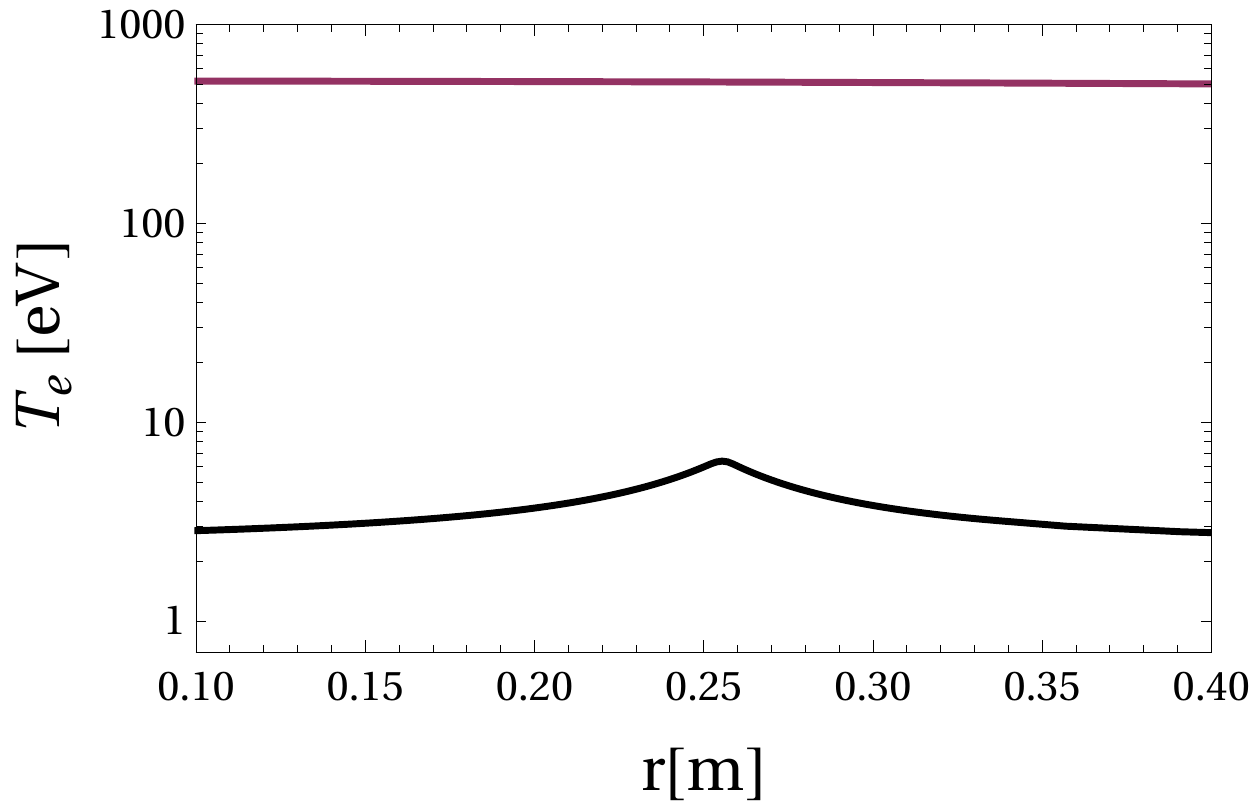}
    \put(-13,55){e)}
    \put(-105,60){ \includegraphics[scale=0.45]{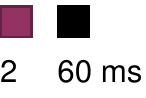}}
    \includegraphics[width=0.33\textwidth]{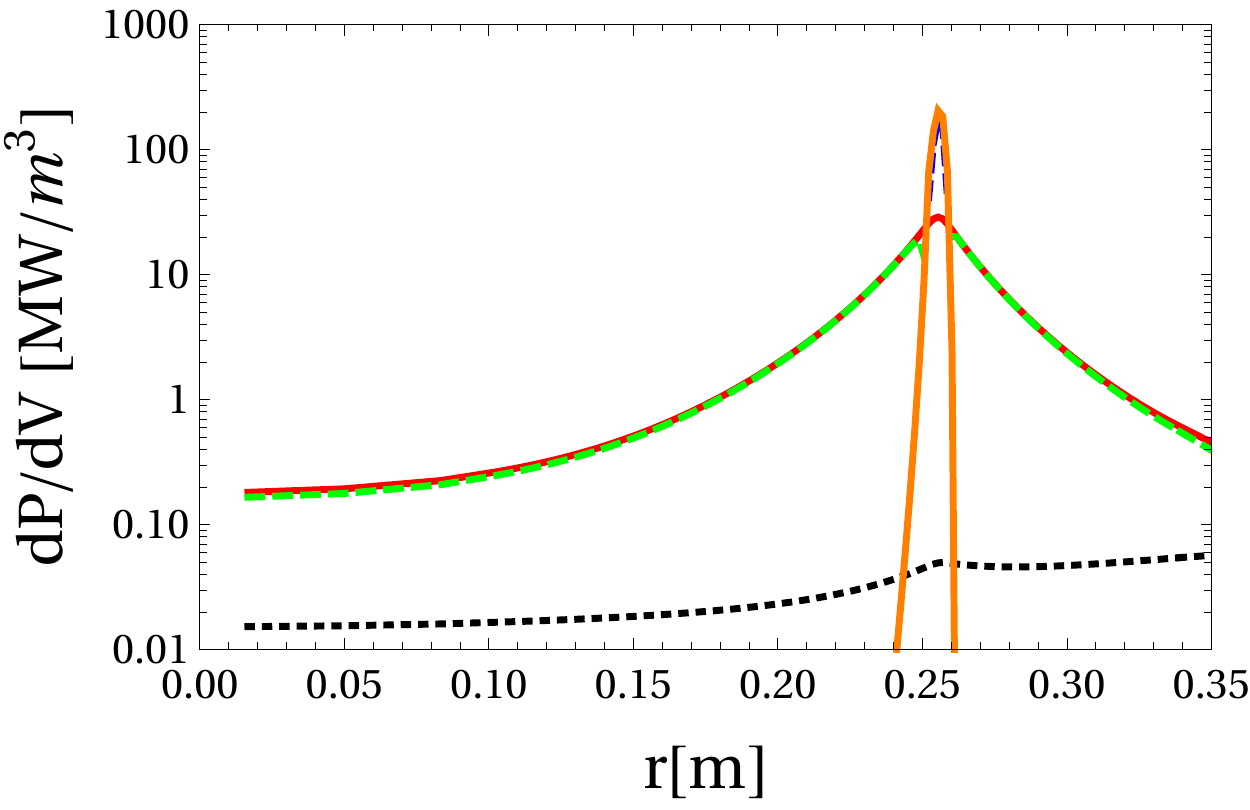}
    \put(-30,30){{\tiny \textcolor{orange}{RE}}}
    \put(-100,20){{\tiny \textcolor{black}{ohmic}}}
    \put(-55,72){{\tiny \textcolor{blue}{transp.~out}}}
    \put(-68,50){{\tiny \textcolor{red}{radiation}}}
    \put(-100,37){{\tiny \textcolor{green}{transp.~in}}}
    \put(-13,55){f)}
    \caption{Dynamics of the \emph{Intact core} case with low (a-c) and high (d-f) remnant heat diffusivity (zoom-in around skin current region; note the logarithmic scales). Panels a and d: Radial profiles of the current density at various time points, with ohmic (solid) and runaway (dashed) components. Panels b and e: Radial profiles of temperature at various time points. Panels c and f: Power balance showing ohmic heating (dotted), heating by runaways (orange solid), radiative losses (red solid), and net heat transport out from (blue dashed) and into (green dashed) a given radial location; taken at $t=60\,\rm ms$. 
    }
    \label{fig:C5d}
\end{figure}

We now consider the possibility of a partial current relaxation, where a range of flux surfaces in the deep core remain intact, while the magnetic field becomes chaotic in most of the plasma up to the edge. As often observed in 3D MHD disruption simulations \citep{Sommariva_2018,Bandaru_2021,Artola_2022}, the deep plasma core tends to be a region where flux surfaces are not completely destroyed, or are broken up only for a very short period within the thermal quench. This may be related to the typically low magnetic shear in the deep core, which does not favor island overlap. 

To model such a scenario, we assume that the magnetic surfaces are broken up in the first $6\,\rm ms$, when we employ a runaway and an electron heat transport corresponding to $\delta B/B  = 3.5\cdot 10^{-3}$, as well as a hyperdiffusivity of $\Lambda = 3\cdot 10^{-2} \, \rm Wb^2 m/s$, as in the case considered previously. However, now the current relaxation and the runaway electron transport are only active in the outer part of the plasma. The transport coefficients, $D_{\rm RE}$ and $\Lambda$, now denoted by $X$, jump between their core and edge values $X_{\rm core}$ and $X_{\rm edge}$ according to the following radial variation 
\begin{equation}
   X= \frac{X_{\rm core}\left\{\Erf\left[\frac{a-r_m}{w}\right] + \Erf\left [\frac{r_m-r}{w}\right] \right\} +X_{\rm edge}\left\{\Erf\left[\frac{r_m}{w}\right] + \Erf\left[\frac{r-r_m}{w}\right] \right\} }{\Erf\left[\frac{a-r_m}{w}\right] + \Erf\left[\frac{r_m}{w}\right] },
   \label{eq:erf}
\end{equation}
with the error function $\Erf$, and here $X_{\rm core}=0$, $X_{\rm edge}$ corresponding to the values given above, a transition radius $r_m=0.3\,\rm m$ and a characteristic transition width of $w=0.01\,\rm m $. Again, for easier comparison with other cases, we keep $D_W$ radially constant.

With intact regions, and associated skin currents, remnant (i.e.~pre-TQ) heat transport has a major role in the dynamics. 
First we consider a lower remnant electron heat transport at $\delta B/B=4\cdot 10^{-4}$. Similarly to Fig.~\ref{fig:C12_IP}c and a, we observe a rapid flattening of the current density throughout most of the plasma, and an associated $I_p$ spike. At the same time, a strongly peaked skin current appears at the boundary of the intact region, shown in the zoomed-in logarithmic plot of Fig.~\ref{fig:C5d}a. This skin current broadens out with time due to electric field diffusion, and turns into a long-lived ohmic current channel, such that the ohmic decay of $I_p$ halts. In the meantime avalanche, slowly but steadily, leads to an increasing runaway current. This runaway current grows in a very narrow layer around $r=0.25\,\rm m$, where $E/E_D$ spikes first during the formation of the skin current. In this region the hot-tail seed generation peaks around $0.5\, \rm ms$, reaching a maximum runaway rate of $2.7 \cdot 10^{17}\,\rm m^{-3}s^{-1}$, which is seven orders of magnitude higher than the maximum of the tritium seed runaway rate that has the second highest value among the seed generation mechanisms. As the seed generation is exponentially sensitive to the electric field, the seed in the skin layer is many orders of magnitude higher than elsewhere.  Since there is no transport of runaways in the intact region, the generated seed does not broaden radially. Consequently, essentially all runaway generation in the intact region happens in this very narrow peak throughout the simulation, even though $E/E_D$ is not the highest in this region after the current relaxation.   

The ohmic current channel in the intact region broadens inward in time, and it locally heats the plasma above $100\,\rm eV$, as seen in Fig.~\ref{fig:C5d}b. Once the runaway current sheet increased to macroscopic values at the outer edge of the ohmic channel, the ohmic current density peak moves inward, along with the temperature peak. It is interesting to consider the inner structure of the hot ohmic channel in terms of power balance, shown in Fig.~\ref{fig:C5d}c, taken at $t=60\,\rm ms$. The main heat source inside the temperature peak is ohmic heating (black dotted line). As the temperature in the middle of the peak is too high for radiative losses to be dominant, ohmic heating is balanced by an outward diffusion of heat there  (blue dashed). Towards the sides of the temperature peak this transported heat is being deposited (dashed green), and as the temperature drops rapidly outward of the peak---along with the ohmic heating---it is the divergence of the outward heat flux that becomes the dominant heat source, balanced by radiative heat losses (solid red) that are more effective below $100\,\rm eV$. We can also see the contribution of the friction of runaway electrons on the bulk to the heating (orange solid), but at this time point this contribution is subdominant to the ohmic heating. 

The hot current channel exhibits strong similarities to the hot current filaments discussed in detail by \citet{Putvinski97}. Such a formation with the extremely large temperature gradients at its edges are likely to be unstable to microinstabilities, thus its existence would likely be of transient nature, if it could be formed at all. Nevertheless, they are acceptable solutions within the model employed here, where the heat diffusivity is simply prescribed. Heat transport driven by microinstabilities tends to be stiff, with fluxes rapidly increasing above a critical gradient; however in this work any excess of instability thresholds is not being monitored. 

To account for the activity of such instabilities, we may simply increase the remnant heat diffusivity, which can essentially change the dynamics of the skin current region. Plots similar to Fig.~\ref{fig:C5d}a-c, but corresponding to a remnant electron heat diffusivity at $\delta B /B = 2 \cdot 10^{-3}$ (compared to $\delta B/B=4\cdot 10^{-4}$) are shown in Fig.~\ref{fig:C5d}d-f. This time the ohmic current quench completes within $60\, \rm ms$, since the formation of a long-lived hot ohmic current channel is inhibited in the presence of the higher heat diffusivity. Clearly the hot current channel does not form, as seen in the $j$  and $T_e$ plots, Fig.~\ref{fig:C5d}d and e. Once the current relaxation is over the electric field also drops to small values, while it still remains significant enough to drive the avalanche, with the runaway current exceeding  $1\, \rm MA$ before $20\,\rm ms$. Most of the runaway current is carried by a narrow current channel where the seed formation was the strongest, similarly to the case with lower heat diffusivity. However, the power balance has an essentially different character to the lower heat diffusivity case; compare Fig.~\ref{fig:C5d}c and f. As the ohmic current have decayed to small values, the contribution of ohmic heating (black dotted) is negligible compared to the heating from  runaway friction (orange solid). This heating is balanced by a transport outward from the very narrow layer where the runaways are localized. Further out the heat deposited (green dashed) is being removed by radiation (red solid). We note that on a longer time scale the lower heat diffusivity case would also develop into a similar state.   

Since no MHD stability is being monitored in these simulations, such a runaway---or ohmic---current layer can develop enormous current densities, so that the total current it carries becomes comparable with the current in the rest of the plasma. Once that happens, or perhaps even earlier, the current sheet would become MHD unstable, which would again increase all transport channels, as well as would lead to a flattening of the current profile. It is plausible that this process can be modeled by an inward expansion of the region with finite $\Lm$. We note that in  simulations with prescribed inward propagating $\Lm$ (not shown here), we observe that the skin current layer does not disappear, instead it moves inward along with the boundary between the intact and chaotic field line regions. Such behavior is indeed observed in 3D MHD simulations, e.g. in Fig.~9 of \citep{Bandaru_2021}. This inward expansion of the chaotic region may proceed until the field becomes chaotic down to the magnetic axis, or the reformation of flux surfaces may stop this process earlier, in which case the skin current layer survives\footnote{As a curiosity, according to MHD simulations of ITER disruptions with JOREK, a sharp current density peak might survive for a while right around the magnetic axis, as seen in Fig.~19 of \citep{Nardon_2021}.}.

\subsection{Intact edge; reverse skin current and runaways}
\label{sec:IE}
After studying the dynamics in the presence of an intact core we now consider the possibility of the edge region remaining intact while the core undergoes a current profile relaxation. This situation may be representative of an internal plasma instability, which can also arise even in non-disruptive plasmas, such as during saw-teeth activity. In addition, it can also be relevant for scenarios exhibiting ``inside-out thermal quench'', which can occur in case of a shell pellet injection \citep{ShellPellet}, but may also happen in shattered pellet injection cases, depending on the detailed dynamics of impurity deposition and the radiative collapse.

The simulation setup is similar to the intact core case considered in Sec.~\ref{sec:IC}, but now the spatial variation of transport coefficients, according to Eq.~(\ref{eq:erf}), uses  a transition radius of $r_m=1.9\,\rm m$, $X_{\rm edge}=0$, and $X_{\rm core}$ corresponding to $\delta B/B  = 3.5\cdot 10^{-3}$ for runaway particle transport, and a hyperdiffusivity of $\Lambda = 3\cdot 10^{-2} \, \rm Wb^2 m/s$. These transport coefficients are active  for $t< 6\,\rm ms$, along with a radially constant electron heat diffusion at the same magnetic perturbation amplitude; the latter is reduced to $\delta B/B  = 4\cdot 10^{-4}$ afterwards.

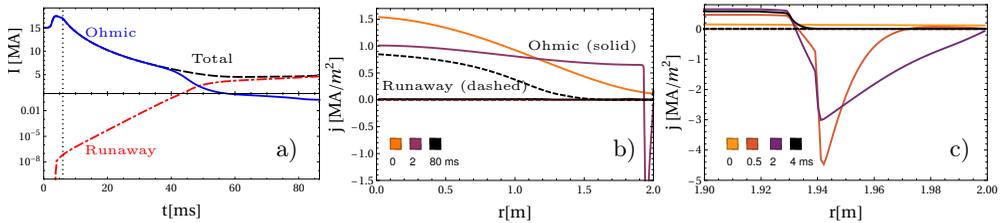
\begin{figure}
    \centering
    \includegraphics[width=0.32\textwidth]{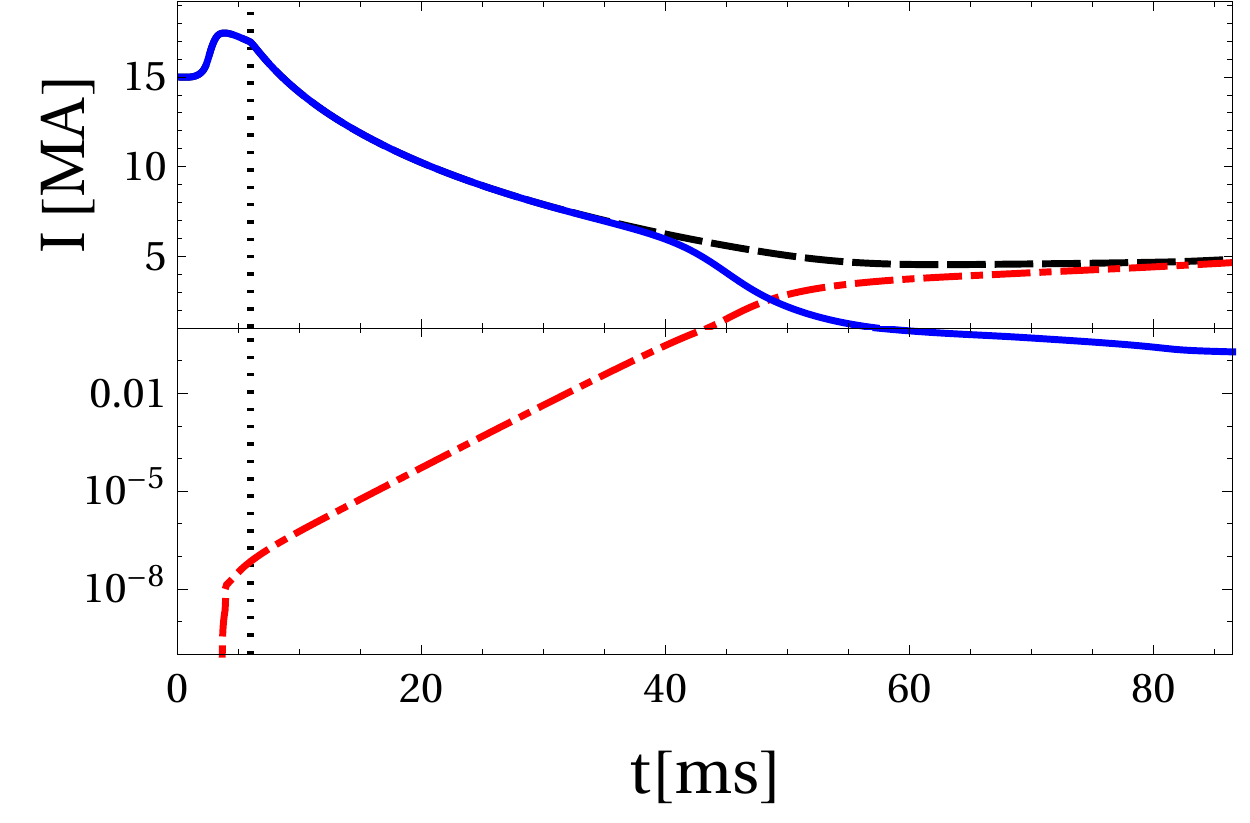}
    \put(-18,25){a)}
    \put(-90,70){\tiny  \textcolor{blue}{Ohmic}}
    \put(-90,26){\tiny  \textcolor{red}{Runaway}}
    \put(-50,60){\tiny  \textcolor{black}{Total}}
    \includegraphics[width=0.33\textwidth]{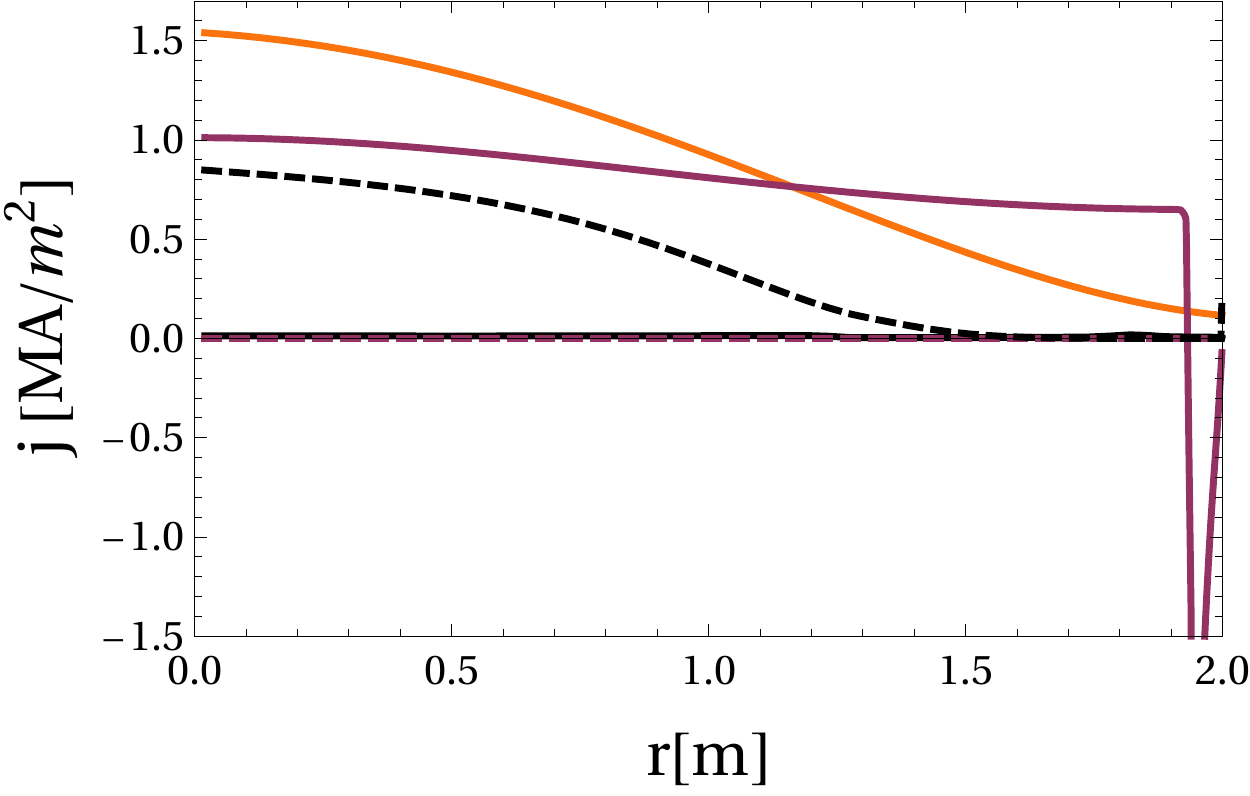}
    \put(-18,25){b)}
    \put(-105,20){ \includegraphics[scale=0.45]{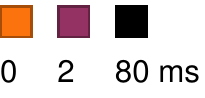}}
    \put(-105,50){\tiny  \textcolor{black}{Runaway (dashed)}}
    \put(-50,65){\tiny  \textcolor{black}{Ohmic (solid)}}
    \includegraphics[width=0.33\textwidth]{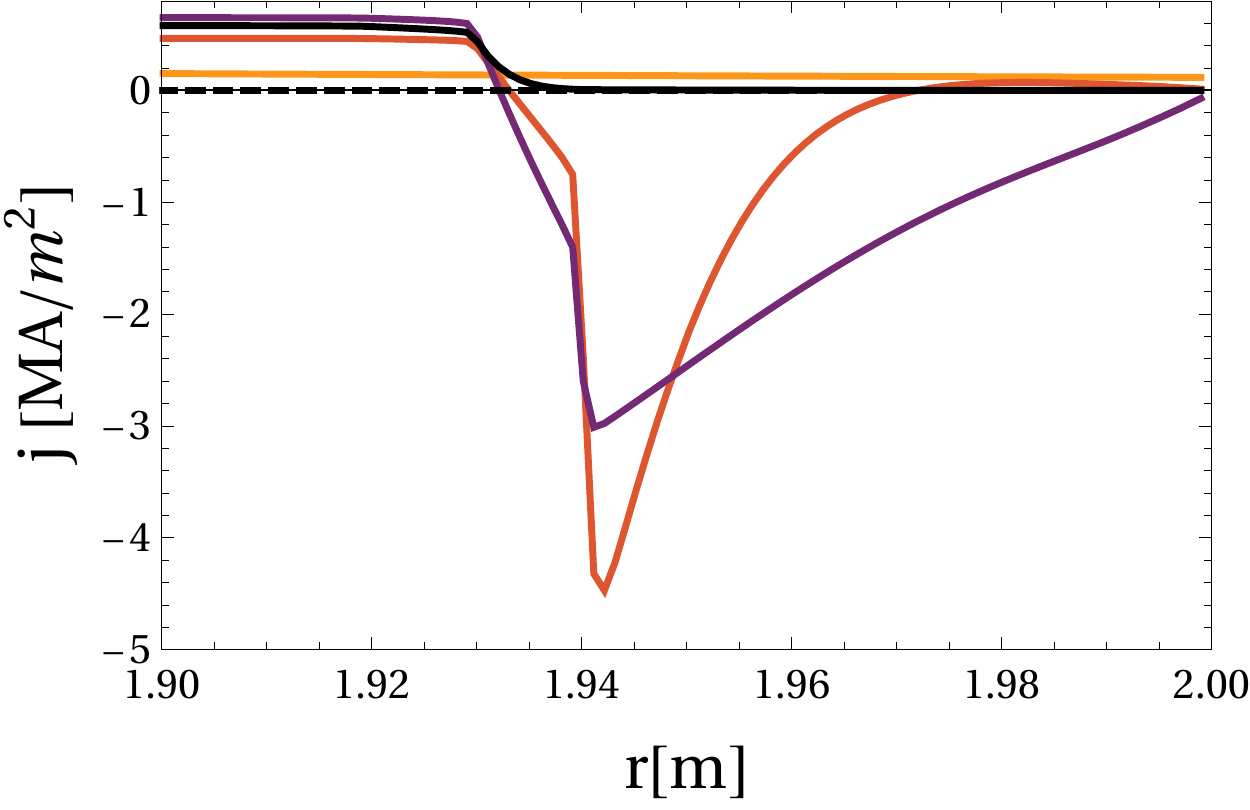}
    \put(-18,25){c)}
    \put(-105,20){ \includegraphics[scale=0.45]{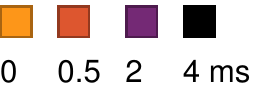}}
    \caption{ Intact edge case yielding a negligible reverse runaway current. Panel a: Time evolution of the total plasma current (dashed) and its ohmic (solid) and runaway (dash-dotted) components; enhanced transport is applied at $t<6\,\rm ms$, indicated by dotted line. Panel b: Current density profiles at various times, with ohmic (solid) and runaway (dashed) components. Panel c: Current density profiles in the skin layer during current relaxation. 
    }
    \label{fig:C7}
\end{figure}

As the current density flattens in the chaotic field region, as seen in Fig.~\ref{fig:C7}b, a strong reverse skin current is generated in the intact edge region. Since the temperature is moderate ($\sim 1\,\rm keV$) in this edge region already in the beginning and it drops below $\sim 50\,\rm eV$ within $2 \,\rm ms$, the resistive diffusion in this region is significant. This leads to a rapid diffusive decay of the reverse skin current; the current density in the skin layer changes direction already at $t=3.5\,\rm ms$. Its evolution during the current relaxation is shown in Fig.~\ref{fig:C7}c. The peak electric field magnitude in the skin layer ($0.26\% E_D$) exceeds that in the rest of the plasma during the course of the entire simulation ($0.13\% E_D$), however it decays too rapidly to lead to a significant reverse runaway current generation; the maximum reverse runaway current density is only $2.5\,\rm mA/m^2$. Nevertheless, the reason for the runaway current curve (dash-dotted) in Fig.~\ref{fig:C7}a to go outside the plotted logarithmic range in the first few milliseconds is the presence of a negative runaway current in the skin layer (reaching only $-2.4\,\rm mA$).

Due to the short magnetic diffusion timescale in the edge, changes in the poloidal magnetic flux during the reconnection is not trapped inside the plasma too long, and an $I_p$ spike is observed already during the time period of the current relaxation, as seen in Fig.~\ref{fig:C7}a (solid line). However, while the $I_p$ spike starts with a clear finite positive $dI_p/dt$  in case there is no intact region in the edge, see e.g.~Fig.~\ref{fig:C12_IP}a, now there is no appreciable change  of $I_p$ in the first $\rm ms$, then the $dI_p/dt$ gradually increases before the $I_p$ spike. Such delay of the $I_p$ spike onset, along with the negative skin current, has already been discussed by \citet{Wesson_1990}, although without the possibility to account for runaway electrons.  

How high values the reverse electric field reaches in the skin layer, and the timescale it lasts, determine whether a significant reverse runaway beam could form. These factors, in turn, depend strongly on the resistive diffusion time, and thus ultimately on the temperature of the skin layer. In the presence of internal instabilities which are localized deeper in the core, thus allowing for a higher temperature, experiments have shown the generation of superthermal electrons in measurable quantities \citep{Klimanov_2007,Kamleitner_2015,MAI_2021}.   
To illustrate that it is conceivable to get significant reverse runaway beams in disruptions with an intact edge region, we return again to the alternative scenario detailed in Appendix~\ref{sec:initailprof}, but now, unlike in the case studied in Fig.~\ref{fig:alternativeEj}, with an intact edge region outside $r_m=1.9\,\rm m$. In this case a fully kinetic simulation is performed, which is practically necessary in case a major reverse runaway population develops during the simulation.

\begin{figure}
    \centering
    \includegraphics[width=0.325\textwidth]{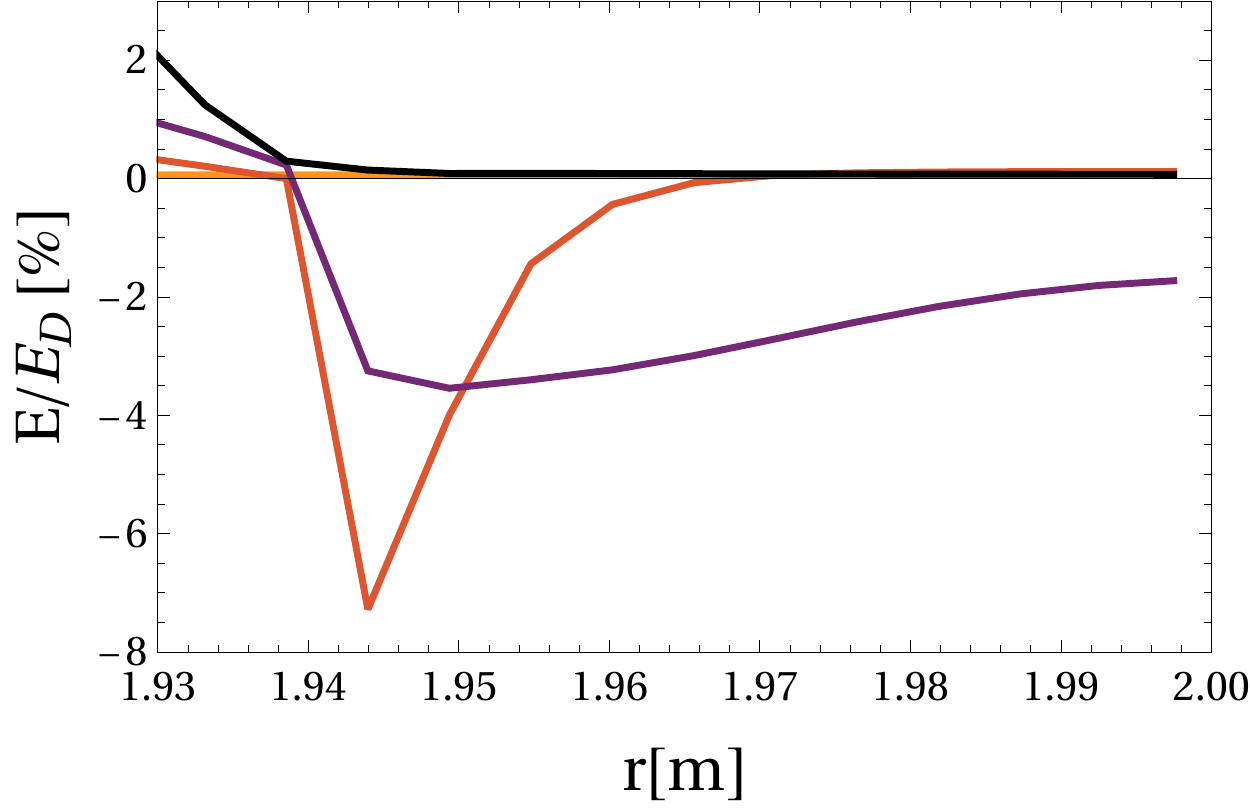}
    \put(-103,20){a)}
    \put(-45,20){ \includegraphics[scale=0.45]{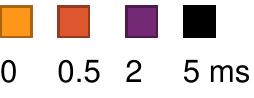}}
    \includegraphics[width=0.33\textwidth]{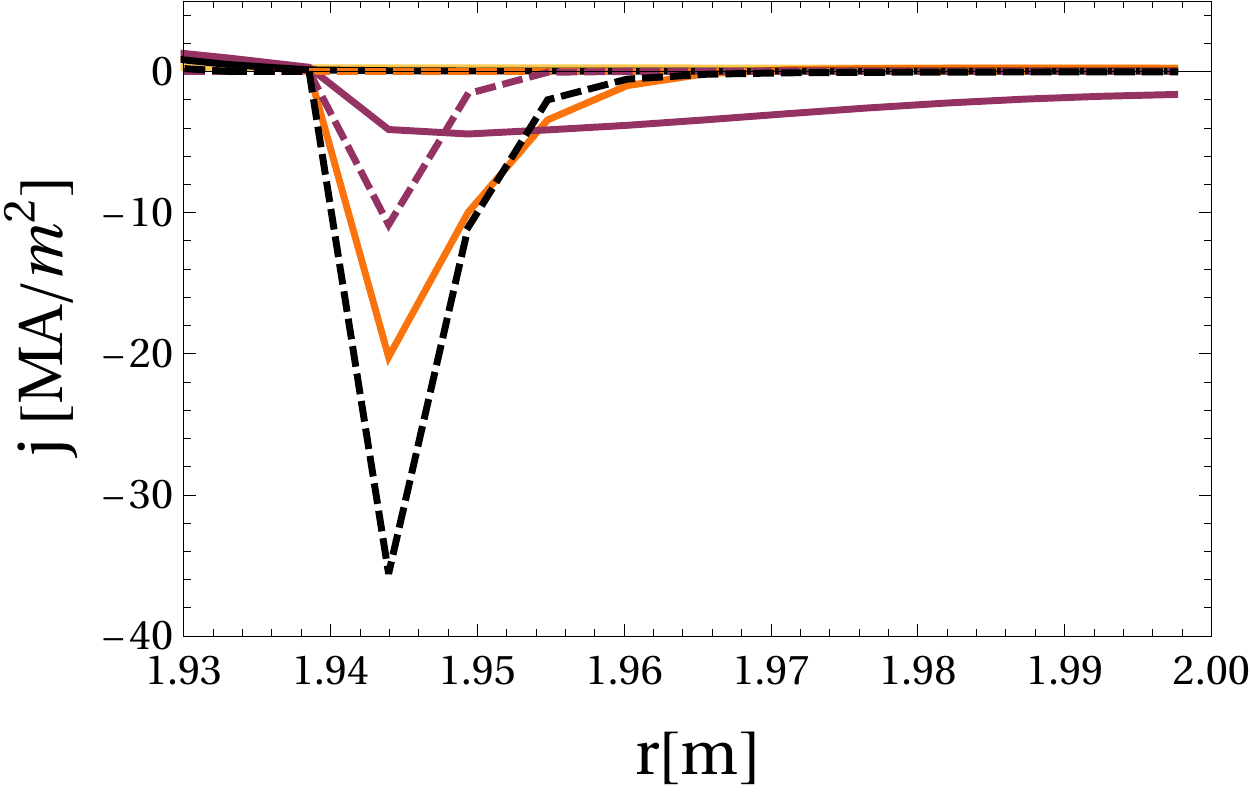}
    \put(-103,20){b)}
    \put(-45,20){ \includegraphics[scale=0.45]{Fig_5a_2.pdf}}
    \put(-80,40){\tiny  \textcolor{black}{Runaway (dashed)}}
    \put(-50,65){\tiny  \textcolor{black}{Ohmic (solid)}}
    \includegraphics[width=0.325\textwidth]{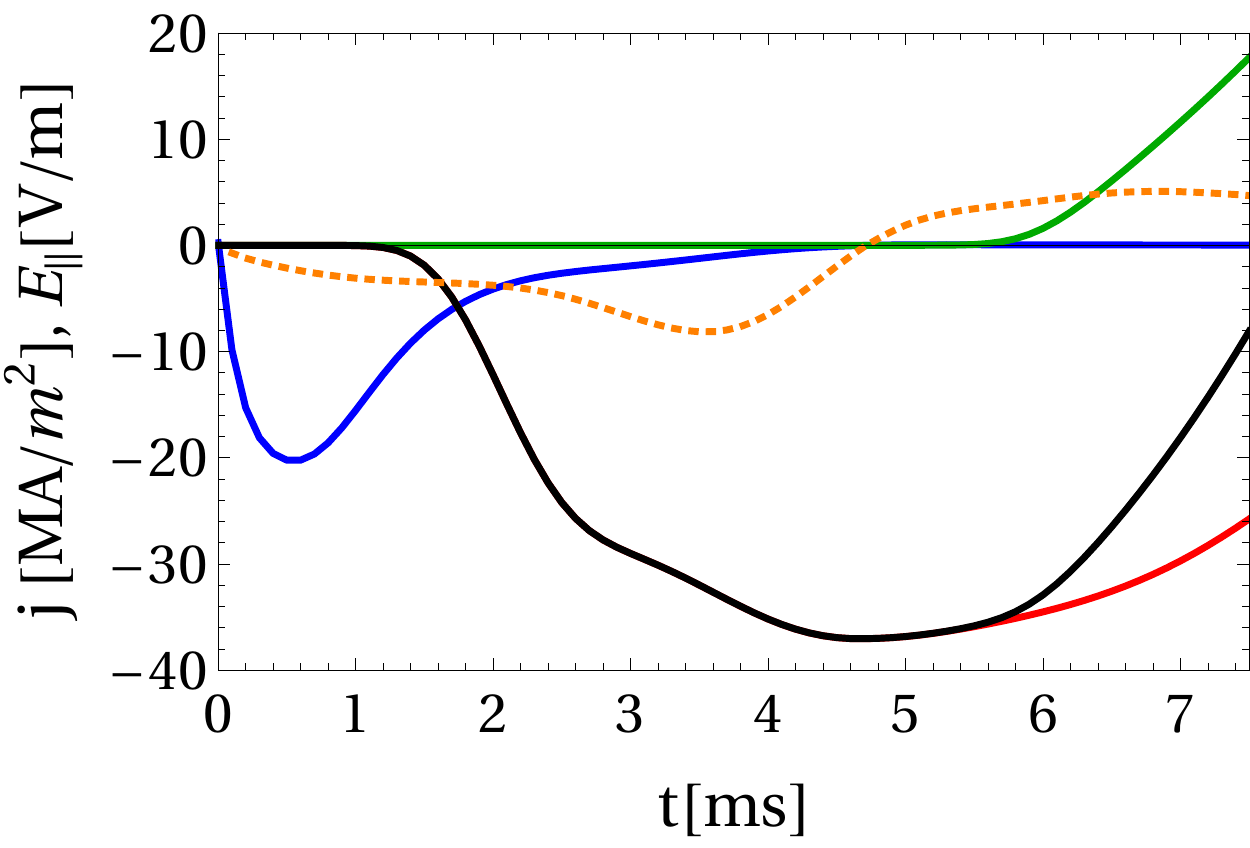}
    \put(-100,20){c)}
    \put(-99,45){\tiny  \textcolor{blue}{$j_{\Omega}$}}
    \put(-14,22){\tiny  \textcolor{red}{$j_{\rm RE}^{-}$}}
    \put(-24,70){\tiny  \textcolor{olive}{$j_{\rm RE}^{+}$}}
    \put(-24,38){\tiny  $j_{\rm RE}$}
    \put(-52,46){\tiny  \textcolor{orange}{$E_\|$}}
    \caption{Intact edge case in an alternative scenario yielding a sizable reverse runaway beam (kinetic simulation). Panel a: Radial profiles of the electric field normalized to the local Dreicer field.  Panel b: Radial profiles of the current density, with ohmic (solid) and runaway (dashed) components.  Panel c: Time evolution of the parallel electric field (orange dotted line) and current density components (solid lines) taken at $r=1.944\,\rm m$. Blue $j_{\Omega}$ -- ohmic, black $j_{\rm RE}$ -- total runaway, red $j_{\rm RE}^{-}$ -- reverse runaway, green $j_{\rm RE}^{+}$ -- forward runaway. }
    \label{fig:IEalternative}
\end{figure}

In this scenario $-E/E_D$ reaches $8.1\%$ compared to $0.2\%$ in the baseline case. This difference translates to a dramatic disparity for the runaway current densities reached; the maximum negative runaway current density reached in the simulation is $36\,\rm MA/m^2$, and the highest negative value of the total runaway current is $3.2\,\rm MA$. The runaway current density is localized to a $\sim 1\,\rm cm$ thin layer, as seen in Fig.~\ref{fig:IEalternative}b. At the location of the highest negative runaway current density, $r=1.944\,\rm m$, the electric field stays negative in the first $4.7\,\rm ms$, and then it changes to the forward direction\footnote{The sign of the electric field is defined such that a positive electric field drives a positive ohmic current that is directed as the initial plasma current.}, see the orange dotted curve in Fig.~\ref{fig:IEalternative}c. 

The electric field first draws a reverse ohmic skin current (blue line in Fig.~\ref{fig:IEalternative}c), which starts to diffusively decay already after $0.5\,\rm ms$. A macroscopic reverse runaway current density is generated a bit after $1\,\rm ms$, which starts to dominate the local current density already before $2\,\rm ms$, and keeps growing as long as the electric field is negative (see red solid curve, overlaid here by the black curve, the total runaway current density). Soon after the electric field becomes positive, a positively directed runaway current component arises (green). Note that the maximum positive electric field at this location reaches only $0.2\%$, which would not be sufficient to generate a macroscopic runaway current under such a short time. However, energetic superthermal electrons which are present due to the \emph{reverse runaway population} can drift over to the positive direction, and turn into an avalanching \emph{forward runaway population}, without first slowing back to the bulk. We note that the avalanche source term we employ takes into account the momentum direction of the runaways that generate new ones by close collisions. For a while the backward and forward runaway populations coexist in the remnant of the skin current region.       

\subsection{Outlook}
\label{sec:discussion}
The analysis presented here is not fully self-consistent as the transport coefficients are prescribed. As we have shown, there is a fair degree of robustness to changes in transport coefficients, in particular to the transport of runaways during the thermal quench, or the remnant electron heat transport after the thermal quench. However, particularly in the presence of skin current regions, it would be desirable to capture the dynamic nature of current and pressure driven instabilities. A fully self-consistent treatment would require nonlinear magnetohydrodynamic simulations with runaway dynamics included. While such numerical models with fluid runaways have become available recently \citep{Bandaru2019,Liu2020} they have their own limitations and represent an incomparably larger computational expense than the approach pursued here. There are various ways to move towards self-consistency, ranging from implementing an automatic enhancement of transport once some specified threshold---in for instance pressure gradient or the tearing parameter $\Delta'$---is exceeded, or utilizing more complex instability criteria and prescriptions based on wisdom gained from nonlinear MHD simulations.  

There is a large degree of freedom in prescribing the spatial and temporal structure of helicity transport and the cases shown here only represent a very limited number of possible scenarios. The approach can be informed by close comparison and even fitting to nonlinear MHD simulations. The machinery to turn simulated magnetic perturbations into transport coefficients and using these in \dream{} simulations is in place \citep{Tinguely_2021}. A natural next step is to extend this workflow to helicity transport. The information gained can be used for exploring larger parameter regions with \dream, which can provide interesting scenarios to be analyzed by higher fidelity models.   

\section{Conclusions}
\label{sec:conclusions}
We have analyzed the runaway electron dynamics in disruptions of ITER-like plasmas in the presence of current relaxation that extends over the entire plasma, or only to limited radial domains. We use a one-dimensional helicity transport model to capture the current relaxation due to a fast magnetic reconnection, and employ the disruption runaway electron modeling framework \dream. During the prescribed reconnection event, besides the magnetic helicity transport, heat and energetic electrons also undergo spatial diffusion, and heat losses are also enhanced by an injected deuterium/neon mixture.     

We find that the current relaxation event reduces the efficiency of runaway generation in the core and increases it towards the edge, thus shifting the center of runaway generation towards the edge. This may lead to disparate outcomes depending on the scenario, including the possibility of reduced core localized runaway generation, or edge localized hollow runaway profiles with larger total runaway current. 

In scenarios where the magnetic field does not become chaotic all the way to the magnetic axis, the strong skin current developing at the boundary of the intact region becomes a dynamically important location. If the remnant electron heat diffusivity after the thermal quench is small, such regions may develop into a hot ohmic current channel, with an interesting internal structure. Not accounting for the development of instabilities, the long-time behavior of these structures is not well represented within the current analysis. Allowing for larger heat diffusivity---that, in scenarios without skin currents, would not have a significant effect---inhibits the development of hot ohmic channels, and instead the skin current region turns into a hot-spot for runaway current generation. Revisiting such situations by the micro and macro stability of the plasma monitored and acted upon, appears to be a fruitful path forward.      

Finally, in intact edge scenarios, where magnetic flux surfaces do not break up all the way to the separatrix, a reverse skin current layer develops. Intact edge scenarios can be relevant in case of internal MHD instabilities, in an "inside-out thermal quench", or if helicity transport is inhibited or strongly reduced towards the edge for some other reason. Not only does the skin current affect the time evolution of the $I_p$ spike, that is reliably measured experimentally, but it can act as a hot-spot for runaway electron evolution, but now in the counter-current direction. Since field diffusion tends to be faster at the less hot plasma edge, the conditions for runaway generation tend to decay more quickly compared to a skin current region in the core. However, kinetic simulation results indicate that it may be possible to develop a non-negligible reverse runaway electron current. Even if a macroscopic reverse electron beam would not have time to develop, once the electric field becomes positive, the reverse runaway seed in the skin region can be accelerated into the forward direction and provide a seed for runaway generation even at electric field strengths that would otherwise be too weak for a significant seed generation. The dynamics of reverse skin current regions is worth further investigation, and can be connected to experiments of superthermal electron observations in the presence of large sawtooth crashes.        

\section*{Acknowledgements} 
The authors are grateful for T\"{u}nde F\"{u}l\"{o}p, Eric Nardon, Allen H.~Boozer and Javier Artola for fruitful discussions. This work was supported by the Swedish Research Council (Dnr.~2018-03911). The work has been carried out within the framework of the EUROfusion Consortium, funded by the European Union via the Euratom Research and Training Programme (Grant Agreement No 101052200 — EUROfusion). Views and opinions expressed are however those of the author(s) only and do not necessarily reflect those of the European Union or the European Commission. Neither the European Union nor the European Commission can be held responsible for them.

\appendix

\section{Initial plasma and magnetic geometry profiles}
\label{sec:initailprof}

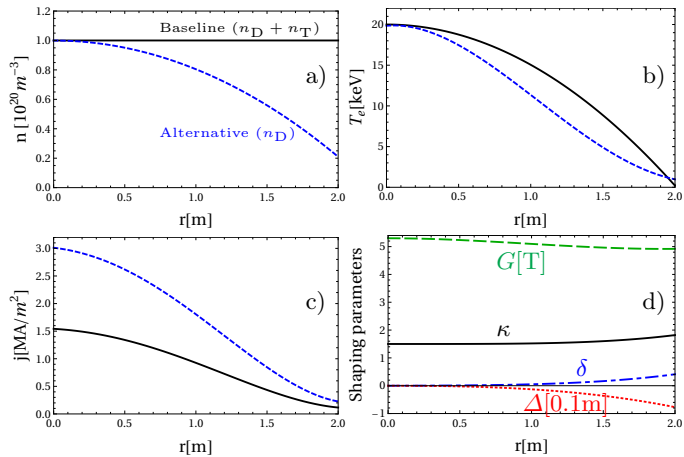
\begin{figure}
    \centering
    \includegraphics[width=0.33\textwidth]{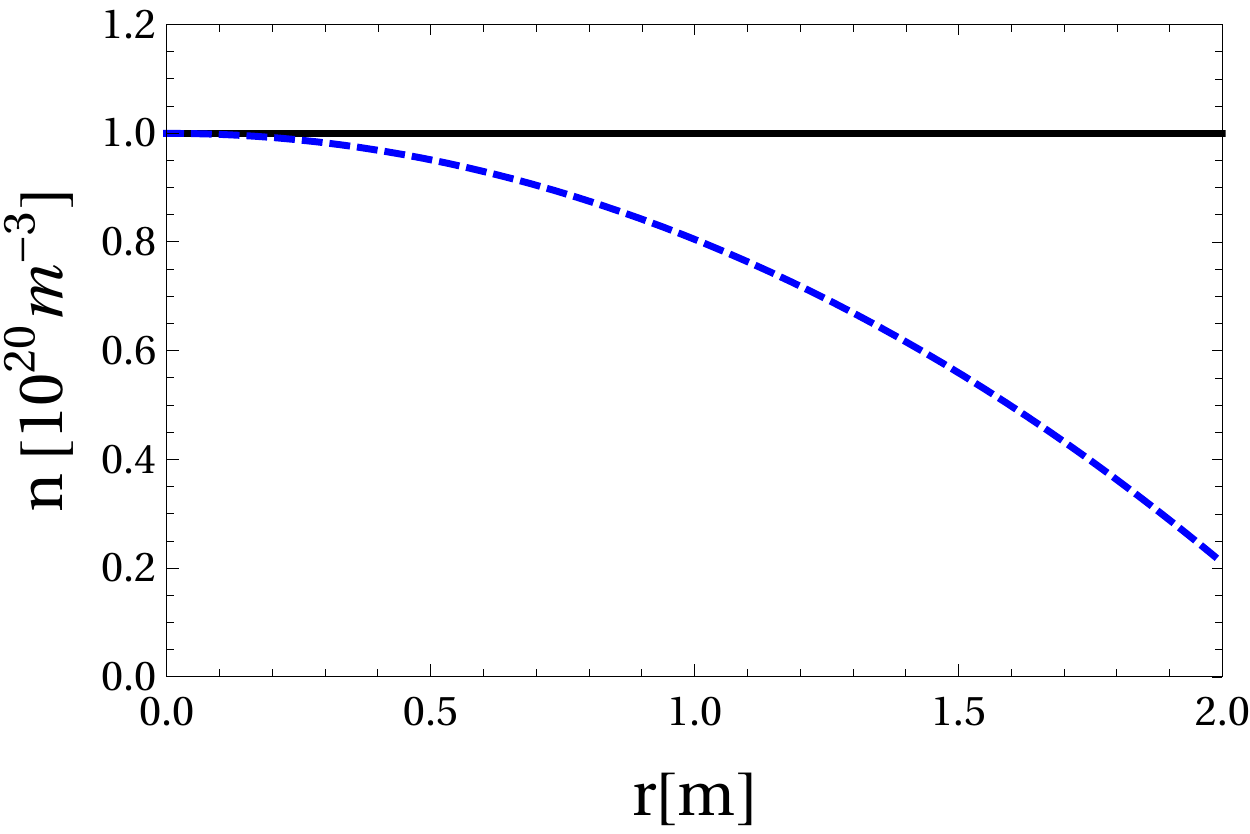}
    \put(-15,55){a)}
    \put(-70,35){\tiny \textcolor{blue}{Alternative ($n_{\rm D}$)}}
    \put(-70,74){\tiny Baseline ($n_{\rm D}+n_{\rm T}$)}
    \includegraphics[width=0.33\textwidth]{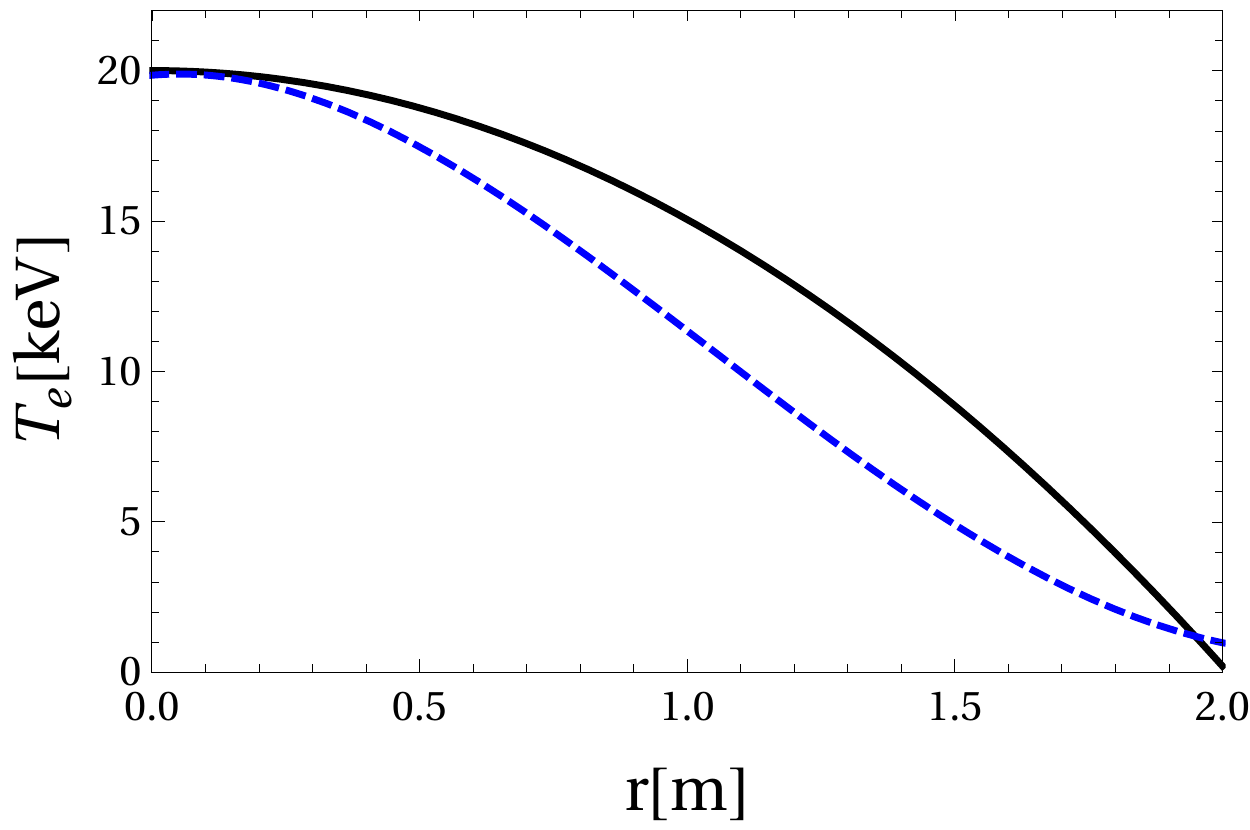}
    \put(-15,55){b)}
    \\
    \includegraphics[width=0.33\textwidth]{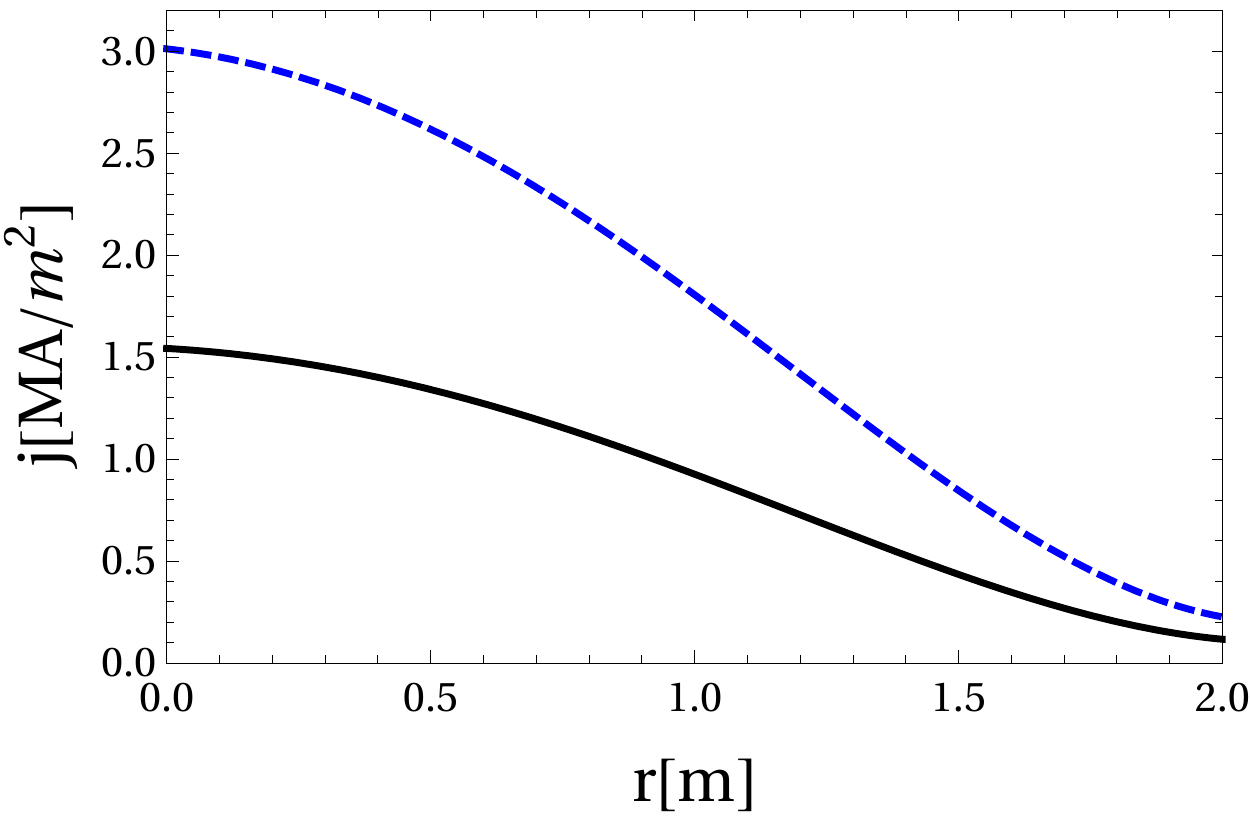}
    \put(-15,55){c)}
    \includegraphics[width=0.33\textwidth]{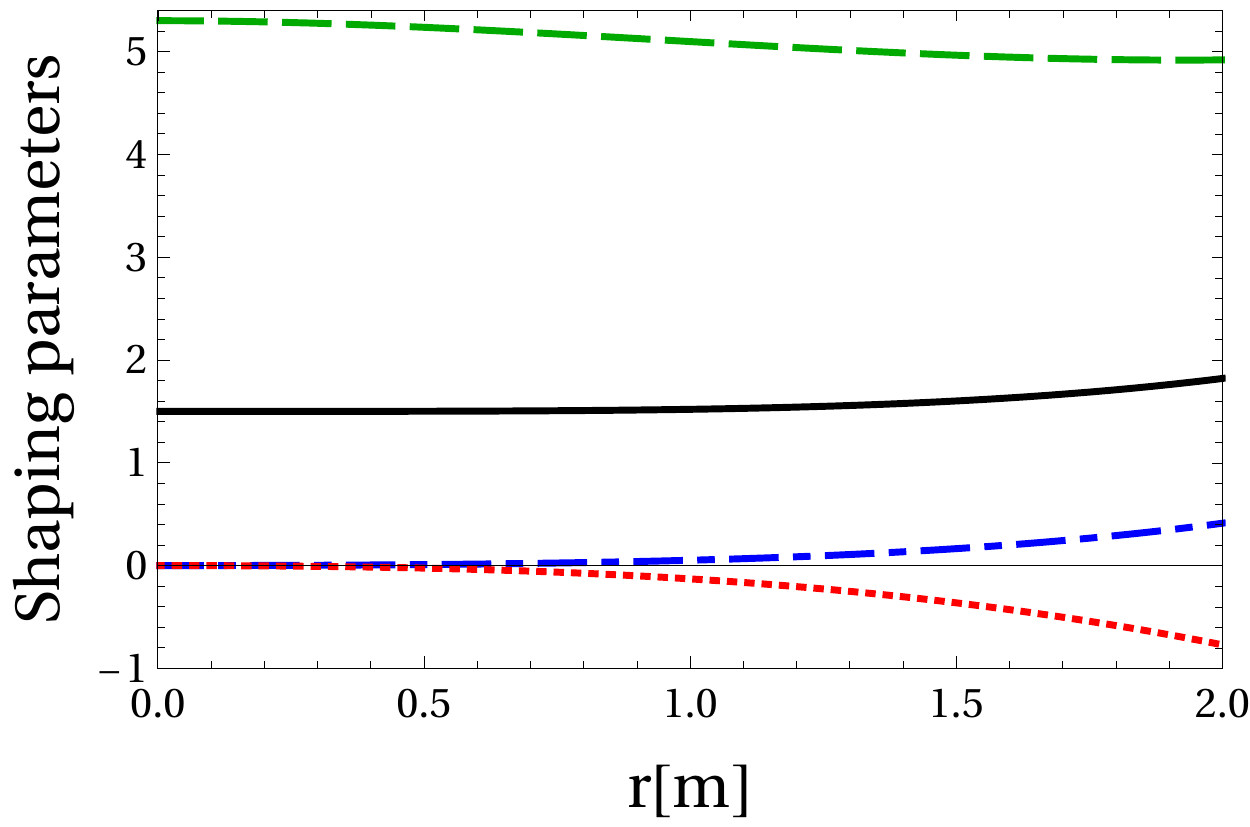}
    \put(-15,55){d)}
    \put(-70,70){\small\textcolor{Green}{$G[{\rm T}]$}}
    \put(-70,45){\small \textcolor{black}{$\kappa$}}
    \put(-40,30){\small \textcolor{blue}{$\delta$}}
    \put(-60,17){\small \textcolor{red}{$\Delta[{\rm 0.1 m}]$}}
    \caption{Panels a-c: Initial plasma profiles in the baseline (solid) and the alternative (dashed) case; a: Total density of hydrogen isotopes present, b: electron temperature, c: current density (the total current in both cases is $15\,\rm MA$, which corresponds to a lower $j$ in the elongated baseline case). Panel d: Shaping parameters in the baseline case.
    }
    \label{fig:profiles}
\end{figure}

The initial plasma parameter profiles are shown in figure~\ref{fig:profiles}a-c in the baseline case (solid curves) and the alternative case (dashed). The current density $j$ is taken at the outboard mid-plane, which is the definition used throughout the paper (and in \dream). In the baseline case the plasma  is shaped with magnetic geometry parameters shown in figure~\ref{fig:profiles}d. The geometric quantity $G$ determines the magnitude of the toroidal magnetic field $B_{\rm t}=|G\nabla\varphi|$, with the toroidal angle $\varphi$; its on-axis value is $B_0=5.3\,\rm T$. Furthermore the elongation $\kappa$ and the triangularity $\delta$ are defined as in the Miller model equilibrium \citep{Miller98}, and the Shafranov shift parameter, $\Delta=R(r)-R_0$, is defined here to be zero on-axis. The magnetic equilibrium is not evolved self-consistently in the simulation, thus these shaping parameters are held fixed.

\textbf{The alternative case} also represents an ITER-sized plasma, but the physics content of the simulation is strongly reduced compared to that of the baseline. 
In this case we use a cylindrical model of the plasma (e.g., all toroidicity effects neglected and there is no plasma shaping) with minor radius $a=2\,\rm m$ and major radius $R_0=6\,\rm m$. A perfectly conducting wall is located at the radius $b=1.35 a$, and the on-axis magnetic field is $B_0=5.3\,\rm T$. The plasma composition is prescribed (charge states are not evolved), and it consists of fully ionized deuterium with density, as shown in \ref{fig:profiles}a, and Ar$^{5+}$ of density $n_{\rm Ar}=0.2 n_{\rm D}$. The temperature evolution is prescribed as
\begin{equation}
  T_e(r,t)=T_f+[T_e(r,t=0)-T_f]\exp(-t/t_{\rm TQ}),
\end{equation}
where the final temperature $T_f=3\,\rm eV$ is radially constant, and the characteristic time of the temperature decay is $t_{\rm TQ}=1\,\rm ms$. This choice of $T_f$ is somewhat arbitrary; it is smaller than typical early post-thermal quench temperatures tend to be ($5$-$10 \,\rm eV$), and it favors a large seed generation. This circumstance is not crucial for the behavior we intend to illustrate with this example (for instance, favoring runaway generation towards the edge). Here, the plasma conductivity is calculated using the collisionless limit of the model of \citep{Redl}. Only Dreicer and avalanche runaway generation mechanisms are active, using the same models as in the baseline case. Runaway electrons are not transported radially in these simulations. 

In the kinetic simulation presented in Fig.~\ref{fig:IEalternative} we set $\Lm=1.5\cdot 10^{-2} \,\rm Wb^2 m/s$ inside of the radius $r_m=1.9\,\rm m$ transitioning to zero outside, over a characteristic distance of $w=5\,\rm mm$. This simulation uses $40$ radial grid cells, $25$ of which are packed around the skin current region. The momentum space of the thermal and superthermal regions are resolved by $25$ cells in $\xi=p_\|/p$, where $p$ is the magnitude of momentum and  $p_\|$ is its component along the magnetic field (taken at the lowest magnetic field point along the orbit). The thermal region extends up to $p=0.07 m_e c$ and is resolved by $60$ cells in $p$, and the superthermal region extends to $p=2 m_e c$ and is resolved by $80$ momentum cells. The runaway region extends up to $p=40 m_e c$ and has $50$ cells in $p$ and $100$ in $\xi$. 

\section{Numerical conservation of magnetic helicity}
\label{helicity}
 Here we demonstrate the conservation of helicity in current relaxation simulations in an example where  $\Lm$ is non-zero in a finite radial domain, and it is zero in the edge and the deep core. In particular,  $\Lm$ makes an error function-like transition between $0$ and $5\cdot 10^{-3}\,\rm Wb^2m/s$ at $r=0.6\,\rm m$ and back to $0$ at $r=1.6\,\rm m$, over characteristic length scales of $2\,\rm mm$ (analogously to Eq.~(\ref{eq:erf}) that only has one such transition). The simulation uses a magnetic geometry and analytically specified plasma parameter profiles as given in Appendix~\ref{sec:initailprof}. The temperature is prescribed as temporally constant, thus, owing to the low resistivity, the total plasma current only decreases by a relative factor of $6\cdot 10^{-4}$ throughout the simulation of $0.01\,\rm s$ duration.     
 
\begin{figure}
	\centering
	\includegraphics[width=0.47\textwidth]{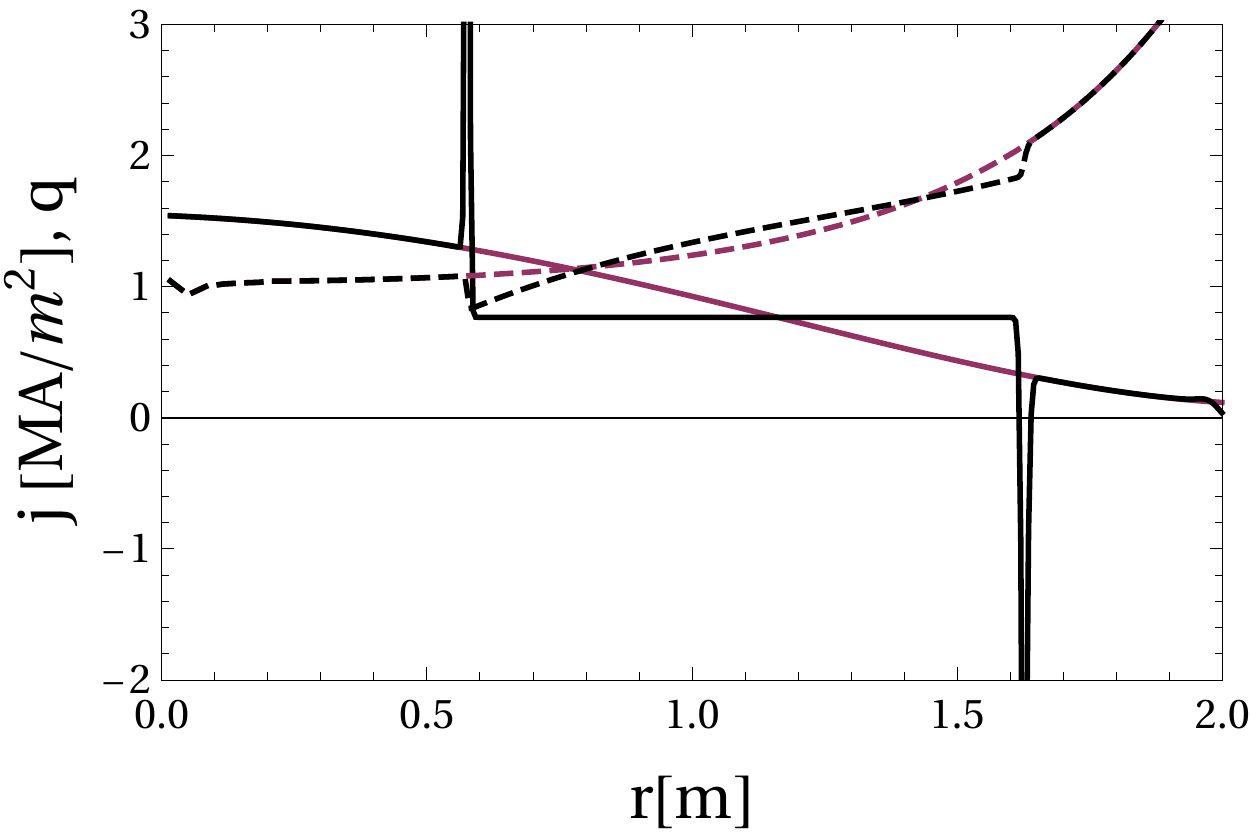}	
	\caption{Variation of current density (solid) and safety factor (dashed) profiles during current relaxation event in a test of helicity conservation, with a finite radial region with non-zero helicity transport. Purple: $t=0$, black: $t=0.01\,\rm s$.}
	\label{fig:helic}
\end{figure}

As seen in Fig.~\ref{fig:helic}, the current density (solid lines) undergoes a complete relaxation in the region of finite $\Lm$, while thin and large skin current densities---positive and negative---form inside of the boundaries of the intact regions (the one in the core reaching $26\,\rm MA/m^2$). Similarly, the $q$ profile also undergoes non-negligible changes (dashed lines). The magnetic helicity, as numerically calculated by the integral
\begin{equation}
H^{\rm M}(t)=-\int_0^a q(r,t) \frac{\partial [\psi_p^2(r,t)]}{\partial r} dr,   
\label{helicity}    
\end{equation}
with poloidal flux $\psi_p$, changes only by a relative amount $3.8\cdot 10^{-5}$ throughout the simulation, even though very sharp current structures---resolved only by a few grid cells---develop in the skin layers. This helicity change is the same order of magnitude as that in a similar simulation without current relaxation ($\Lm=0$). Thus the size of the helicity change we observe is consistent with being caused by the finite resistivity of the plasma. The radial resolution used is $400$ grid cells for $0.1<r/a<0.95$, and $50$ cells distributed in the remaining radial domain. 



\bibliographystyle{jpp}
\bibliography{bibliography}

\end{document}